\documentclass[aps,10pt,prd,groupedaddress,nofootinbib,notitlepage,eqsecnum,preprintnumbers]{revtex4-2}
\usepackage[utf8]{inputenc}
\usepackage{hyperref}
\usepackage{xcolor}
\usepackage{graphicx}
\usepackage{amsmath,amssymb}
\usepackage{bm}
\usepackage[shortlabels]{enumitem}
\begin{document}
\preprint{YITP-21-31}
\title{Cosmology of strongly interacting fermions in the early universe}

\author{\textsc{Guillem Dom\`enech$^{a}$}}
    \email{{domenech}@{pd.infn.it}}
\author{\textsc{Misao Sasaki$^{b,c,d}$}}
    \email{{misao.sasaki}@{ipmu.jp}}

\affiliation{$^{a}$ \small{INFN Sezione di Padova, I-35131 Padova, Italy}\\      $^{b}$\small{Kavli Institute for the Physics and Mathematics of the Universe (WPI), Chiba 277-8583, Japan}\\
      $^{c}$\small{Center for Gravitational Physics, Yukawa Institute for Theoretical Physics, Kyoto University, Kyoto 606-8502, Japan}\\
      $^{d}$\small{Leung Center for Cosmology and Particle Astrophysics, National Taiwan University, 
      Taipei 10617, Taiwan}}

\begin{abstract}
In view of growing interest in long range scalar forces in the early universe to generate primordial black holes, we study in detail the general relativistic formulation of a Fermi gas interacting with a scalar field in cosmology. Our main finding is that the Yukawa interaction leads to a solution where the scalar field oscillates around zero fermion mass and all energy densities decay as radiation. On one hand, we show that if the Fermi gas starts relativistic, it could stay relativistic. On the other hand, if the fermions are initially non-relativistic, they remain non-relativistic for all practical purposes. We find that in both cases the energy density of the fermions and the scalar field decays as radiation. In the non-relativistic case, this is due to an oscillating and decaying effective mass. Such background dynamics questions whether there is a substantial enhancement of the fermion density fluctuations in the non-relativistic case. Our work can be easily extended to more general field dependent fermion mass. The analysis of the cosmological perturbations will be presented in a follow-up work.
\end{abstract}
 \maketitle

\section{Introduction \label{sec:Intro}}

The direct detection of gravitational waves by LIGO \cite{Abbott:2016nmj} and the image of the shadow in M87 galaxy's center by the EHT \cite{Akiyama:2019cqa} give very strong evidence of the presence of black holes, ranging from ten to billion solar masses. While these black holes might have an astrophysical origin, it is entirely possible that a fair portion have formed in the early universe. The existence of primordial black holes (PBHs) would be immediately confirmed if a black hole with less than a solar mass is ever observed. In general, PBHs have a very rich phenomenology (e.g. see \cite{Sasaki:2018dmp,Carr:2020xqk} for a review): They might be a substantial fraction (if not all) of the dark matter \cite{Carr:2020xqk} (and references therein), they could be responsible for some of the LIGO/VIRGO gravitational waves (GWs) events \cite{Bird:2016dcv,Sasaki:2016jop,Wong:2020yig} and the microlensing events by planet-mass objects found by OGLE \cite{2017Natur.548..183M,Niikura:2019kqi}. PBH might also be the seeds of supermassive black holes \cite{Kawasaki:2012kn,Carr:2018rid}.

The most studied mechanism for the formation of PBHs dates back to the works of Hawking and Carr \cite{Hawking:1971ei,Carr:1974nx}. It involves the collapse of large primordial fluctuations, which must have been generated during inflation. However, this is not the only possibility. For example, PBHs could form by collapse of vacuum bubbles, nucleated during inflation, after the universe reheats \cite{Garriga:2015fdk}. They could come from first order phase transitions \cite{Crawford:1982yz,Kodama:1982sf} and the collapse of Q-balls \cite{Cotner:2016cvr,Cotner:2019ykd}. They might also be the result of long range interactions stronger than gravity \cite{Amendola:2017xhl}, although the end state of such collapse is still debated \cite{Savastano:2019zpr,Flores:2020drq}. Thus, even the absence of PBHs constrains the physics of the early universe. 

Long range scalar forces have been considered in the late universe as an interaction between dark matter and dark energy \cite{Amendola:1999er}, which impacts structure formation \cite{Farrar:2003uw}. The effect of coupling a scalar field, in the mentioned case is the quintessence \cite{Wetterich:1987fm}, to a non-relativistic fluid yields an effective time dependent mass for the non-relativistic particles \cite{Amendola:1999er}. This mechanism has also been used as a coupling between quintessence dark energy, a SU(2) triplet scalar and neutrinos, leading to time dependent neutrino masses \cite{Amendola:2007yx,Wetterich:2007kr} by the Seesaw II mechanism \cite{Magg:1980ut,Schechter:1980gr}. Only recently it has been applied to PBH formation in the early universe \cite{Amendola:2017xhl,Flores:2020drq}. However, as already noticed in the context of growing neutrinos in \cite{Bento:2009xa,Casas:2016duf,MohseniSadjadi:2017pne} the non-relativistic description is incomplete. This is clear from the following argument. The scalar force is long range if the mass of the scalar field is small compared to the scale of interest, in the PBH formation case it is the horizon scale. This massless scalar field evolves towards minimizing the effective mass of the non-relativistic particle. Eventually, the effective mass becomes so small that the non-relativistic approximation of a gas of particles is no longer valid. Thus, it is important to have a fully relativistic formulation of the system with analytical background solutions. The background solutions shall be then used to compute the evolution of perturbations and to precisely check if, given adiabatic (or isocurvature) initial conditions, the non-relativistic number density has time to grow. 

Let us briefly mention the type of coupling used in the literature. On one hand, Ref.~\cite{Amendola:2017xhl} considers a dilatonic (exponential) coupling of the scalar field to the non-relativistic particles. This choice is motivated by non-minimal couplings to gravity or string theory \cite{Fujii:2003pa}, first proposed by Jordan \cite{Jordan:1959eg} and Brans \& Dicke \cite{Brans:1961sx}. The dilatonic coupling is particularly useful because the system made of radiation, non-relativistic matter and the scalar field has an attractor solution where all energy densities behave as radiation (so called scaling solution). This simplifies considerably calculations. However, the scaling solution should change once the relativistic regime is achieved. On the other hand, Ref.~\cite{Flores:2020drq} takes a Yukawa (linear) coupling to non-relativistic fermions. The Yukawa coupling is strongly motivated by the standard model of particle physics with the Higgs interactions \cite{Higgs:1964pj,Englert:1964et}. However, the background solutions are only known in the small field regime. In this work, we provide a fully relativistic formulation of the long range scalar interaction, assuming thermal and chemical equilibrium. We show that even in the Yukawa case, the system reaches a similar scaling solution around a vanishing effective mass. Our methodology can be applied to any coupling which renders the particle masses field dependent. The study of the cosmological perturbations will be provided in a follow-up work.

This paper is organized as follows. In Sec.~\ref{sec:SIF} we review the formulation of a Fermi gas in an expanding universe. We consider a field dependent mass a provide the consistent set of general relativistic equations in the grand canonical ensemble. In Sec.~\ref{sec:deg}, we study the case of a degenerate Fermi gas, which can be treated fully analytically. In there, we show that the system reaches a scaling solution. In Sec.~\ref{sec:nondeg}, we extend the analysis to the non-degenerate case and find that it shares similarities with the non-degenerate case. We summarize and discuss further implications of our work in Sec.~\ref{sec:conclusions}. Details of the calculations can be found in the appendices. Throughout this paper we work in natural units, i.e. $\hbar=c=1$.

\section{Fermions with a Yukawa interaction in curved spacetime \label{sec:SIF}}

We are interested in the cosmology of massive fermions, say $\psi$, which have a Yukawa interaction with a scalar field $\varphi$. The microscopic Lagrangian for Dirac fermions with the Yukawa interaction is given by \cite{Srednicki:1019751}
\begin{align}\label{eq:dirac}
{\cal L}_{\psi}=\bar\psi i\Gamma^\mu D_\mu\psi-m_{\psi}\bar\psi\psi-y\varphi\bar\psi\psi\,,
\end{align}
where $\Gamma^\mu$ are the Dirac matrices in curved spacetime, $D_\mu$ is the covariant derivative which contains interaction with gauge fields, $m_\psi$ is the bare mass and $y$ is the strength of the Yukawa coupling. Due to the $U(1)$ gauge symmetry, we have net particle number conservation.
Now, assuming that the scalar field is homogeneous, as usual in cosmological situations, we conclude that in the macroscopic picture, and in the mean field approximation, the only effect of the Yukawa interaction is a field dependent effective mass. It should be noted that depending on the scalar field value, the effective mass might become negative. Nevertheless, this is not a problem for the theory as one may change the sign of the mass term by a chiral transformation, that is $\psi\to\Gamma^5\psi$ with $\Gamma^5$ being the generalization of the Dirac matrix $\gamma^5$ in curved spacetime. Thus, we shall define the effective mass to be always positive definite \cite{Srednicki:1019751} and given by
\begin{align}\label{eq:meff}
m_{\rm eff}=|m_\psi+y\varphi|\,.
\end{align}
However, the sign relative sign between $m_\psi$ and $y\varphi$ might be very important when considering the change of $m_{\rm eff}$ with $\varphi$. To keep track of the sign, we introduce the following quantity:
\begin{align}\label{eq:sigma}
\sigma\equiv{\rm sign}[m_\psi+y\varphi]\,. 
\end{align}
It should be noted that since the scalar field does not violate fermion number the chemical potential is independent of the scalar field. We also assume that the massive fermions and their associated gauge bosons are part of a dark sector, invisible to the standard model of particle physics. 

Before going into the details, let us note that in the exact massless scalar field limit the bare mass $m_\psi$ might be redefined by a scalar field redefinition.\footnote{For example, this is clear if as in Ref.~\cite{Farrar:2003uw} we write the coupling as $y(\varphi-\varphi_*)$ where $\varphi_*\equiv -m_\psi/y$.} This is because in the absence of a potential for $\varphi$, the system has a shift symmetry. Namely, the system is invariant under $\varphi\to\varphi+C$ together with $m_\psi\to m_\psi-C$. Thus, any constant value of $\varphi$ might be thought of as a redefinition of the bare mass $m_\psi$. This also means that, in the massless scalar field regime, the system will evolve towards minimizing the energy and, therefore, towards $m_{\rm eff}\to 0$ irrespective of the initial values of $m_\psi$ and $\varphi$.

\subsection{Brief review of thermodynamics\label{subsec:reviewthermo}}

In cosmology, rather than the microscopic description \eqref{eq:dirac}, we have a gas of fermions described by the Fermi-Dirac statistics. Thus, we can gain considerable intuition starting with a thermodynamical point of view. From the first law of thermodynamics, we have that the internal energy of the Fermi gas satisfies \cite{Fetter}
\begin{align}\label{eq:microcanonical}
dE_\psi=TdS_\psi-P_\psi dV+\mu dN_\psi+Y_\psi d\varphi\,,
\end{align}
where $S_\psi$ is the entropy, $P_\psi$ the pressure, $V$ the volume, $N_\psi$ the net particle number and we are for the moment treating $\varphi$ as an external force. The dynamics of the scalar field $\varphi$ are given by the Klein-Gordon equation which we derive later. We use the subscript $\psi$ to indicate that they refer to the Fermi gas. Considering the extra contribution from the scalar force, the energy conservation reads
\begin{align}\label{eq:energyconservation1}
\frac{dE_\psi}{dt}+P_\psi\frac{dV}{dt}=Y_\psi\frac{d\varphi}{dt}\,,
\end{align}
where the coupling $Y_\psi$ is derived according to
\begin{align}\label{eq:Y1}
Y_\psi\equiv\left(\frac{\partial E_\psi}{\partial \varphi}\right)_{S_\psi,N_\psi,V}\,.
\end{align}
This means that if we know how the energy depends on the scalar field, we automatically have the form of the coupling. Also it is important to note that from Eq.~\eqref{eq:microcanonical}, entropy conservation $dS_\psi/dt=0$ follows from the energy conservation \eqref{eq:energyconservation1} and number particle conservation $dN_\psi/dt=0$. Using the scaling properties of the extrinsic/intrinsic variables we also have that \cite{Fetter}
\begin{align}
E_\psi=TS_\psi-P_\psi V +\mu N_\psi\,,
\end{align}
where we used that the homogeneous scalar field $\varphi$ remains the same under a rescaling.

The universe is well described by an homogeneous and isotropic flat Friedmann-Lema{\^i}tre-Robertson-Walker (FLRW) metric, explicitly given by
\begin{align}
ds^2=-dt^2+a^2dx^2\,,
\end{align}
where $a$ is the scale factor. We see that in the FLRW universe the volume scales as $V\propto a^3$. It then follows that the energy density conservation reads
\begin{align}\label{eq:energyconservation2}
\dot \rho_\psi+3H(\rho_\psi+P_\psi)=\dot\varphi\left(\frac{\partial \rho_\psi}{\partial \varphi}\right)_{s_\psi,n_\psi}\,,
\end{align}
where $\dot\,\equiv \frac{\partial}{\partial t}$ and $H=\dot a/a$.
Note that we moved to intrinsic variables defined by $\rho_\psi\equiv E_\psi/V$, $s_\psi\equiv S_\psi/V$ and $n_\psi\equiv N_\psi/V$. All quantities in \eqref{eq:energyconservation2} should be regarded as a function of $n_\psi$, $s_\psi$ and $\varphi$.

In cosmological situations, however, the gas of fermions is well described by the grand canonical ensemble. This is because each Hubble patch is free to exchange particles and energy with neighboring regions. Further assuming that the Fermi gas is in thermal and chemical equilibrium, which would definitely be the case if the expansion is slow enough (or interactions are strong enough), we have a Fermi gas with constant temperature $T$ and chemical potential $\mu$. Note that due to chemical equilibrium articles and antiparticles have opposite chemical potential. The grand canonical potential is defined by a Legendre transform of the internal energy given by \cite{Fetter}
\begin{align}
\Omega=E_\psi-TS_\psi-\mu N_\psi\,.
\end{align}
In this case, we have that the differential reads
\begin{align}\label{eq:domega}
d\Omega=-S_\psi dT-N_\psi d\mu-P_\psi dV+Y_\psi d\varphi\,,
\end{align}
where the fundamental variables are now $V$, $T$, $\mu$ and $\varphi$. Using the scaling properties one finds that \cite{Fetter}
\begin{align}\label{eq:pomega}
\Omega=-P_\psi V.
\end{align}
Using Eqs.~\eqref{eq:domega} and \eqref{eq:pomega} we see that in the grand canonical ensemble the coupling term is determined by
\begin{align}\label{eq:Y2}
Y_\psi= \left(\frac{\partial \Omega_\psi}{\partial \varphi}\right)_{\mu,T,V}=-V\left(\frac{\partial P_\psi}{\partial \varphi}\right)_{\mu,T,V}\,.
\end{align}
For a detailed relation between the quantities in the grand canonical ensemble and the microcanonical ensemble we refer the reader to the appendix of Ref.~\cite{Floerchinger:2015efa}. This time, the energy conservation \eqref{eq:energyconservation2} in the grand canonical ensemble reads
\begin{align}\label{eq:energyconservation3}
\dot\rho_\psi+3H(\rho_\psi+P_\psi)=-\dot\varphi\left(\frac{\partial P_\psi}{\partial \varphi}\right)_{\mu,T}\,,
\end{align}
where each variable must be regarded as a function of $T$, $\mu$ and $\varphi$. Note the difference with Eq.~\eqref{eq:energyconservation2}, where in the microcanonical ensemble the coupling is proportional to the derivative of the energy density. We shall use description \eqref{eq:energyconservation3} when dealing with the full set of Einstein equations.

In thermal and chemical equilibrium, the number and energy densities and pressure for a Fermi gas with mass $m_{\rm eff}$, chemical potential $\mu$ and temperature ${T}$ in the grand canonical ensemble are given by
\begin{align}\label{eq:thermo}
n_\psi&=\frac{2}{(2\pi)^3a^3}\int d^3p \,\left(f(\mathbf{p},m_{\rm eff},\mu)-f(\mathbf{p},m_{\rm eff},-\mu)\right)\,,\\
\rho_\psi&=\frac{2}{(2\pi)^3a^3}\int d^3p \,E(\mathbf{p},m_{\rm eff})\left(f(\mathbf{p},m_{\rm eff},\mu)+f(\mathbf{p},m_{\rm eff},-\mu)\right)\label{eq:rho}\,,\\
P_\psi&=\frac{2}{(2\pi)^3a^5}\int d^3p \,\frac{p^2}{3E(\mathbf{p},m_{\rm eff})}\left(f(\mathbf{p},m_{\rm eff},\mu)+f(\mathbf{p},m_{\rm eff},-\mu)\right)\label{eq:p}\,,
\end{align}
where $p$ is the comoving momentum and the factor $2$ comes from the degrees of freedom of the spin-$1/2$ particles. The Fermi Dirac distribution is given by
\begin{align}
f(\mathbf{p},m_{\rm eff},\mu)=\left({e^{\frac{E(\mathbf{p},m_{\rm eff})-\mu}{{T}}}+1}\right)^{-1}\quad{\rm with}\quad E(\mathbf{p},m_{\rm eff})=\sqrt{\frac{p^2}{a^2}+m_{\rm eff}^2}\,.
\end{align}
Using that $P_\psi$ is proportional to the grand canonical potential, we have the following thermodynamical relations
\begin{align}\label{eq:relationsbg}
n_\psi=\left(\frac{\partial P_\psi}{\partial\mu}\right)_{T,\varphi}\quad{,}\quad s_\psi=\left(\frac{\partial P_\psi}{\partial T}\right)_{\mu,\varphi}=\frac{\rho_\psi+p_\psi-\mu n_\psi}{T}\,,
\quad{\rm and}\quad
\left(\frac{\partial P_\psi}{\partial \varphi}\right)_{\mu,T}=-y\sigma\frac{\rho_\psi-3P_\psi}{m_{\rm eff}}\,.
\end{align}
The entropy $s_\psi$ as well as the number density $n_\psi$ are conserved. These and other useful relations can be found in Appendix~\ref{app:relations}.

Before going to the next subsection, it is important to note that in this paper we assume that interactions with the fermions are such that thermal and chemical equilibrium is maintained and so we can use Eqs.~\eqref{eq:thermo}-\eqref{eq:p}. However, when the Fermi gas has a finite temperature one should specify the interactions and check the equilibrium distribution through the collisionless Boltzmann equations (for a discussion in the context of neutrinos see Ref.~\cite{Lesgourgues:1519137}), specially for the transition regions. For this reason, when dealing with the finite temperature case we only study the relativistic and non-relativistic limits to next to leading order.

\subsection{Equations of motion\label{subsec:fermigasexpanding}}

Let us present the equations of motion of the joint system composed of gravity, radiation, fermions and the scalar field. First, the Einstein Equations are given by
\begin{align}\label{eq:EE}
M_{\rm pl}^2 G_{\mu\nu}=\partial_\mu\varphi\partial_\nu\varphi-\frac{1}{2}g_{\mu\nu}\left(\partial_\alpha\varphi\partial^\alpha\varphi+M^2\varphi^2\right)+T_{\psi,\mu\nu}+T_{R,\mu\nu}\,,
\end{align}
where $M_{\rm pl}^2=1/(8\pi G)$, the subscript $R$ refers to Radiation and $T_{Q,\mu\nu}$ with $Q=\{R,\psi\}$ is the energy-momentum tensor of a perfectly fluid, explicitly given by
\begin{align}\label{eq:Trmunu}
T_{Q,\mu\nu}=\left(\rho_Q+P_Q\right)u_{Q,\mu}u_{Q,\nu}+P_Qg_{\mu\nu}\,.
\end{align}
In Eq.~\eqref{eq:Trmunu}, $u_{Q,\mu}$ are the fluid's 4-velocity. Note that we are assuming that the fermions are in thermal and chemical equilibrium and, thus, are well approximated by a perfect fluid. Second, the energy conservation of the Fermi gas \eqref{eq:energyconservation3} reads
\begin{align}\label{eq:energycons}
u_\psi^\mu\nabla_\mu\rho_\psi+\left(\rho_\psi+P_\psi\right)\nabla_\mu u_\psi^\mu+\left(\frac{\partial P_\psi}{\partial\varphi}\right)_{T,\mu}u_\psi^\mu\nabla_\mu\varphi=0\,.
\end{align}
From the Bianchi identities and the conservation of the energy momentum tensor for radiation, we find that the Klein-Gordon equation is given by
\begin{align}\label{eq:KG}
\nabla_\nu\nabla^\nu\varphi-M^2\varphi+\left(\frac{\partial P_\psi}{\partial\varphi}\right)_{T,\mu}=0\,.
\end{align}
Similarly, we obtain the equation for the velocity as
\begin{align}\label{eq:velocity}
&\left(\rho_\psi+p_\psi\right)u_\psi^\nu\nabla_\nu u_{\psi,\mu}+\left(\delta_{\mu}^\nu+u_{\psi,\mu}u^\nu_{\psi}\right)\left({\nabla_\nu p_\psi}-\left(\frac{\partial P_\psi}{\partial\varphi}\right)_{T,\mu}\nabla_\nu\varphi\right)=0\,.
\end{align}
These equations are supplemented by the number density conservation
\begin{align}
\nabla_\mu(n_\psi u_\psi^\mu)=0\,,
\end{align}
from which, together with energy conservation, entropy conservation follows, that is $\nabla_\mu(s_\psi u_\psi^\mu)=0$.

Now that we have the relativistic equations of motion, we shall derive the background equations of motion. The energy conservation and Klein-Gordon equations respectively read
\begin{align}\label{eq:energyconservationbg}
\dot\rho_\psi+3H(\rho_\psi+P_\psi)-y\sigma\dot\varphi\frac{\rho_\psi-3P_\psi}{m_{\rm eff}}=0\,,
\end{align}
and
\begin{align}\label{eq:KLEIN}
\ddot\varphi+3H\dot\varphi+M^2\varphi+y\sigma\frac{\rho_\psi-3P_\psi}{m_{\rm eff}}=0\,,
\end{align}
where we already used Eq.~\eqref{eq:relationsbg}. The time dependence of the chemical potential and the temperature are determined by the number and energy density conservation. Explicitly, their equations of motion are given by
\begin{align}\label{eq:Tmu1}
\left(\frac{\partial \rho_\psi}{\partial T}\right)_{\mu,\varphi}\dot T+\left(\frac{\partial \rho_\psi}{\partial \mu}\right)_{T,\varphi}\dot\mu+\left(\frac{\partial \rho_\psi}{\partial \varphi}\right)_{T,\mu}\dot\varphi+3H(\rho_\psi+P_\psi)=y\sigma\dot\varphi\frac{\rho_\psi-3P_\psi}{m_{\rm eff}}\,.
\end{align}
and
\begin{align}\label{eq:Tmu2}
\left(\frac{\partial n_\psi}{\partial T}\right)_{\mu,\varphi}\dot T+\left(\frac{\partial n_\psi}{\partial \mu}\right)_{T,\varphi}\dot\mu+\left(\frac{\partial n_\psi}{\partial \varphi}\right)_{T,\mu}\dot\varphi+3Hn_\psi=0\,.
\end{align}
It should be noted that the coupling term in the energy conservation \eqref{eq:energyconservationbg} and Klein-Gordon equation \eqref{eq:KLEIN} is proportional to the trace of the Fermi gas energy momentum tensor \eqref{eq:Trmunu}, as it is usually used \cite{Amendola:1999er,Amendola:2017xhl}. However, we emphasize here that the consistent way to couple the fluids at the level of the equations of motion is through $\left({\partial P_\psi}/{\partial\varphi}\right)_{T,\mu}$ if one uses the grand canonical ensemble. This is in contrast with the naive expectation from the microcanonical ensemble which one uses $\left({\partial \rho_\psi}/{\partial\varphi}\right)_{s_\psi,n_\psi}$.

From now on and otherwise stated, we shall assume for simplicity that radiation completely dominates the universe. Then the Friedmann equations yield
\begin{align}
a(t)=a_i\left(\frac{t}{t_i}\right)^{1/2}\quad{\rm and}\quad H=\frac{1}{2t}\,,
\end{align}
where $t_i$ is an arbitrary pivot time deep inside radiation domination. Any subscript $i$ in the text refers to a quantity evaluated at $t=t_i$. In this approximation, the coupled dynamics of the fermions and the scalar field are obtain by solving Eqs.~\eqref{eq:KLEIN}, \eqref{eq:Tmu1} and \eqref{eq:Tmu2} simultaneously.  In the next sections, we derive analytical formulas in limiting cases of interest. We first deal with a degenerate Fermi gas and then turn to the non-degenerate case. In general situations though it may only be solved numerically.

\section{Degenerate Fermi gas\label{sec:deg}}
In this section we consider the case of a degenerate Fermi gas where all calculation can be performed analytically. In the degenerate case, the fermions are extremely cold\footnote{To avoid any confusions, it should be noted that the fact that the fluid is cold, in the sense that $T\to 0$, is not related with whether the fluid is relativistic or non-relativistic. In the case of fermions, they can be cold and relativistic. By increasing the number density, one also populates the high energy states (due to the Pauli exclusion principle). In other words, fermions may have a large degeneracy pressure.} and decoupled from the thermal bath of the radiation fluid. They only interact through the Yukawa interaction with the scalar field. Explicitly, we take the strict limit ${T}\to0$ and set the chemical potential to be the Fermi energy $\mu=E_{F}(m_{\rm eff})$ for $E<E_F$. In this way, the number density of states will be populated until $E=E_F$. Integrating the number density given by Eq.~\eqref{eq:thermo} we find
\begin{align}
E_F^2=m_{\rm eff}^2+N_\psi^{2/3}\,,
\end{align}
where we defined for later convenience
\begin{align}
N_\psi\equiv{3\pi^2n_\psi}\,.
\end{align}
This means that we can compute all the quantities in terms of $m_{\rm eff}$ and $N_\psi$. In particular, the energy density and pressure of the Fermi gas read
\begin{align}
\rho_\psi&=\frac{1}{8\pi^2}\left(\left(2N_\psi+m^2_{\rm eff}N_\psi^{1/3}\right)\sqrt{m_{\rm eff}^2+N_\psi^{2/3}}-{m_{\rm eff}^4}{\rm arcsinh}\left[\frac{N_\psi^{1/3}}{m_{\rm eff}}\right]\right)\,,\\
P_\psi&=\frac{1}{8\pi^2}\left(\left(\frac{2}{3}N_\psi-m^2_{\rm eff}N_\psi^{1/3}\right)\sqrt{m_{\rm eff}^2+N_\psi^{2/3}}+{m_{\rm eff}^4}{\rm arcsinh}\left[\frac{N_\psi^{1/3}}{m_{\rm eff}}\right]\right)\,.
\end{align}

We can further classify the degenerate Fermi gas by their equation of state parameter $w_\psi\equiv P_\psi/\rho_\psi$ into relativistic $w_\psi=1/3$ and non-relativistic $w_\psi\ll1$. This classification respectively corresponds to the cases $m_{\rm eff}\ll N_\psi^{1/3}$ and $m_{\rm eff}\gg N_\psi^{1/3}$. In this limits we explicitly find that the equation of state reads
\begin{align}
1-3w_\psi\approx \left\{
\begin{aligned}
&2\frac{m_{\rm eff}^2}{N_\psi^{2/3}} & m_{\rm eff}\ll N_\psi^{1/3}\\
&1+\frac{3}{5}\frac{N_{\psi}^{2/3}}{m_{\rm eff}^2} & m_{\rm eff}\gg N_\psi^{1/3}
\end{aligned}
\right.\,.
\end{align}
Similarly, the source term is proportional to
\begin{align}
\left(\frac{\partial P_\psi}{\partial \varphi}\right)_{\mu}\approx -y\sigma \left\{
\begin{aligned}
&\frac{1}{2\pi^2}m_{\rm eff}N_\psi^{2/3} &  m_{\rm eff}\ll N_\psi^{1/3}\\
&n_\psi & m_{\rm eff}\gg N_\psi^{1/3}
\end{aligned}
\right.\,.
\end{align}

\subsection{Background dynamics in a radiation dominated universe \label{subsec:bg}}

In this subsection, we derive analytical solutions in the two limiting cases and compare them with the full numerical calculations. First, we note that in the degenerate case, the conservation of energy \eqref{eq:energyconservationbg} automatically yields the conservation of number density
\begin{align}\label{eq:NPSI}
\dot N_\psi+3HN_\psi=0\,.
\end{align}
This is consistent with the fact that $T=0$ and $\mu=E_F$, so the dynamics of $\mu$ are already determined. Solving Eq.~\eqref{eq:NPSI}, we shall use that $N_\psi\propto a^{-3}$. Now, the Klein Gordon equation \eqref{eq:KG} yields
\begin{align}\label{eq:KG2}
\ddot\varphi+3H\dot\varphi+M^2\varphi+\frac{y}{m_{\psi}+y\varphi}\left(\rho_\psi-3P_\psi\right)=0\,,
\end{align}
where we used that $\sigma m_{\rm eff}=m_{\psi}+y\varphi$. To simplify the discussion, we shall neglect for the moment the scalar field mass $M$. We recover it later when discussing the final cosmology of the system. We shall also assume that initially we have $\varphi=0$. As explained below Eq.~\eqref{eq:sigma},  any constant term to $\varphi$ can be reabsorbed in a redefinition of the bare mass $m_\psi$. Thus, without loss of generality we initially set $\varphi=0$.

For clarity, we shall distinguish four situations according to different limits of the source term to the Klein-Gordon equation:
\begin{enumerate}[label={\rm(}\alph*{\rm)},font=\itshape]
\item Relativistic fermion, $m_{\rm eff}\ll N_\psi^{1/3}$, and small scalar field, $y\varphi\ll m_\psi$. \label{sit:1}
\item Relativistic fermion, $m_{\rm eff}\ll N_\psi^{1/3}$, and saturated scalar field, $m_\psi+y\varphi\sim 0$. \label{sit:2}
\item Non-relativistic fermion, $m_{\rm eff}\gg N_\psi^{1/3}$, and small scalar field, $y\varphi\ll m_\psi$. \label{sit:3}
\item Non-relativistic fermion, $m_{\rm eff}\gg N_\psi^{1/3}$, and saturated scalar field, $m_\psi+y\varphi\sim 0$. \label{sit:4}
\end{enumerate}

Before we study in detail these cases separately, it proves useful to introduce the following definitions:
\begin{align}\label{eq:definitions}
\beta\equiv\frac{yM_{\rm pl}}{m_\psi}\,,\quad \chi\equiv\beta H_it\,,\quad 
\phi\equiv\frac{y \varphi}{m_\psi}\,,\quad x\equiv\frac{m_{\rm eff}}{N_\psi^{1/3}}\,.
\end{align}
Note that in Eq.~\eqref{eq:definitions}, $\beta$ represents the strength of the Yukawa coupling. In the situations we are interested we have that $\beta\gg1$. This is also the general case unless $m_\psi\sim M_{\rm pl}$ and/or $y\ll1$. Furthermore, note that $x$ determines whether the Fermi gas is relativistic ($x\ll1$) or non-relativistic ($x\gg1$). We also define for convenience
\begin{align}\label{eq:omegabar}
\bar\Omega_{\psi,\rm r}\equiv\frac{1}{4\pi^2}\frac{N_\psi^{4/3}}{3M_{\rm pl}^2H_i^2}\quad{\rm and}\quad
 \bar\Omega_{\psi,\rm nr}\equiv\frac{1}{3\pi^2}\frac{m_{\rm eff} N_\psi}{3M_{\rm pl}^2H_i^2}\,,
\end{align}
which respectively correspond to the energy density fraction of the fermion fluid in the exact relativistic and non-relativistic limits. Note that the subscripts ``r'' and ``nr'' are used throughout the text to denote the relativistic and non-relativistic limits. The upper bar is to differentiate them from the exact energy density fraction, namely
\begin{align}\label{eq:densityfraction}
\Omega_\psi\equiv\frac{\rho_\psi}{3M_{\rm pl}^2H^2}\,.
\end{align}

With the above new definitions, the scalar field equation takes a simpler form given by
\begin{align}\label{eq:redefinedeq}
\frac{d^2\phi}{d\chi^2}+\frac{3}{2\chi}\frac{d\phi}{d\chi}+\frac{3}{1+\phi}\bar\Omega_{\psi,{\rm r}/{\rm nr}}f_{{\rm r}/{\rm nr}}(x)=0\,,
\end{align}
where
\begin{align}
f_{\rm r}(x)\equiv2x^2\left(\sqrt{1+x^2}-x^2\sinh^{-1}(1/x)\right)\,,\quad
f_{\rm nr}(x)\equiv \frac{3}{4x}f_{\rm r}(x)\,.
\end{align}
In Eq.~\eqref{eq:redefinedeq} we express Eq.~\eqref{eq:KG2} in terms of the two definitions given in \eqref{eq:omegabar}. This will be useful when distinguishing between the situations \ref{sit:1}-\ref{sit:4}. Similarly, for the density fraction we have that
\begin{align}
\Omega_\psi=\bar\Omega_{\psi,{\rm r}/{\rm nr}}F_{{\rm r}/{\rm nr}}(x)
\end{align}
where
\begin{align}
F_{\rm r}(x)\equiv\frac{1}{2}\left((2+x^2)\sqrt{1+x^2}-x^4\sinh^{-1}(1/x)\right)\,,\quad
F_{\rm nr}(x)\equiv \frac{3}{4x}F_{\rm r}(x)\,.
\end{align}

\subsection{Relativistic regime \label{sec:degrel}}

We start by noting that from Eq.~\eqref{eq:KG2} it may seem that while in the exact relativistic limit, that is when $\rho_\psi=3P_\psi$, the scalar field stays constant. Unless we initially have $m_{\rm eff}=\dot m_{\rm eff}=0$, this is in general not the case. When $m_{\rm eff}\ll N_\psi^{1/3}$, the Klein-Gordon equation \eqref{eq:KG2} reads
\begin{align}\label{eq:KGREL}
\frac{d^2\phi}{d\chi^2}+\frac{3}{2\chi}\frac{d\phi}{d\chi}+6x_i^2\bar\Omega_{\psi,{\rm r},i}\frac{\chi_i}{\chi}=0\,,
\end{align}
where we already used that initially $\phi\ll 1$ and $x_i\ll1$. The particular solution to Eq.~\eqref{eq:KGREL} is given by
\begin{align}\label{eq:solsit1}
\phi_{\rm (a)}\approx -\beta^2 x_i^2\bar\Omega_{\psi,{\rm r},i}\frac{\chi}{\chi_i}\approx -\beta^2 x^2\bar\Omega_{\psi,{\rm r},i}\,,
\end{align}
where we used that $\chi_i=\beta/2$ and the subscript $\rm (a)$ is to emphasize that this solution is valid in situation \ref{sit:1}. Therefore we see that the scalar field grows and it is negative due to the attractive interaction with the fermions. Since the field monotonically grows, there may be a moment when $\phi\to-1$ and $m_{\rm eff}=|1+\phi|\to0$. This is the saturated scalar field case that we study below.

\begin{figure}
\centering
\includegraphics[width=0.5\columnwidth]{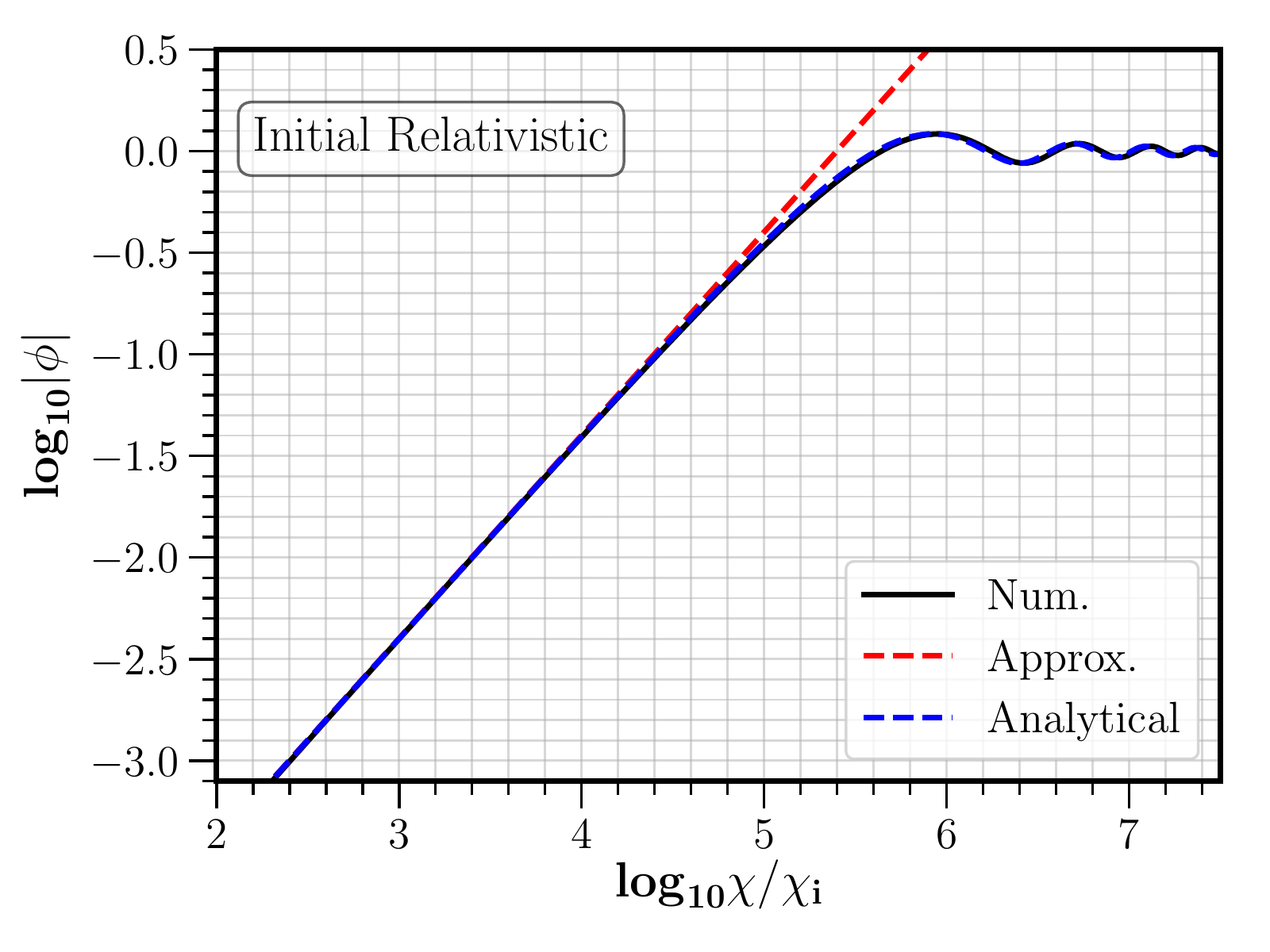}
\caption{Scalar field evolution when the fermion fluid is initially relativistic and stays relativistic. In the figure we chose $\beta=200$, $x_i=10^{-3}$ and $\bar\Omega_{\psi,{\rm r},i}=10^{-4}$. The solid black line is the result of numerical integration while the dashed blue and red lines are the analytical approximations given in the main text.\label{fig:rel}}
\end{figure}

\begin{figure}
\centering
\includegraphics[width=0.5\columnwidth]{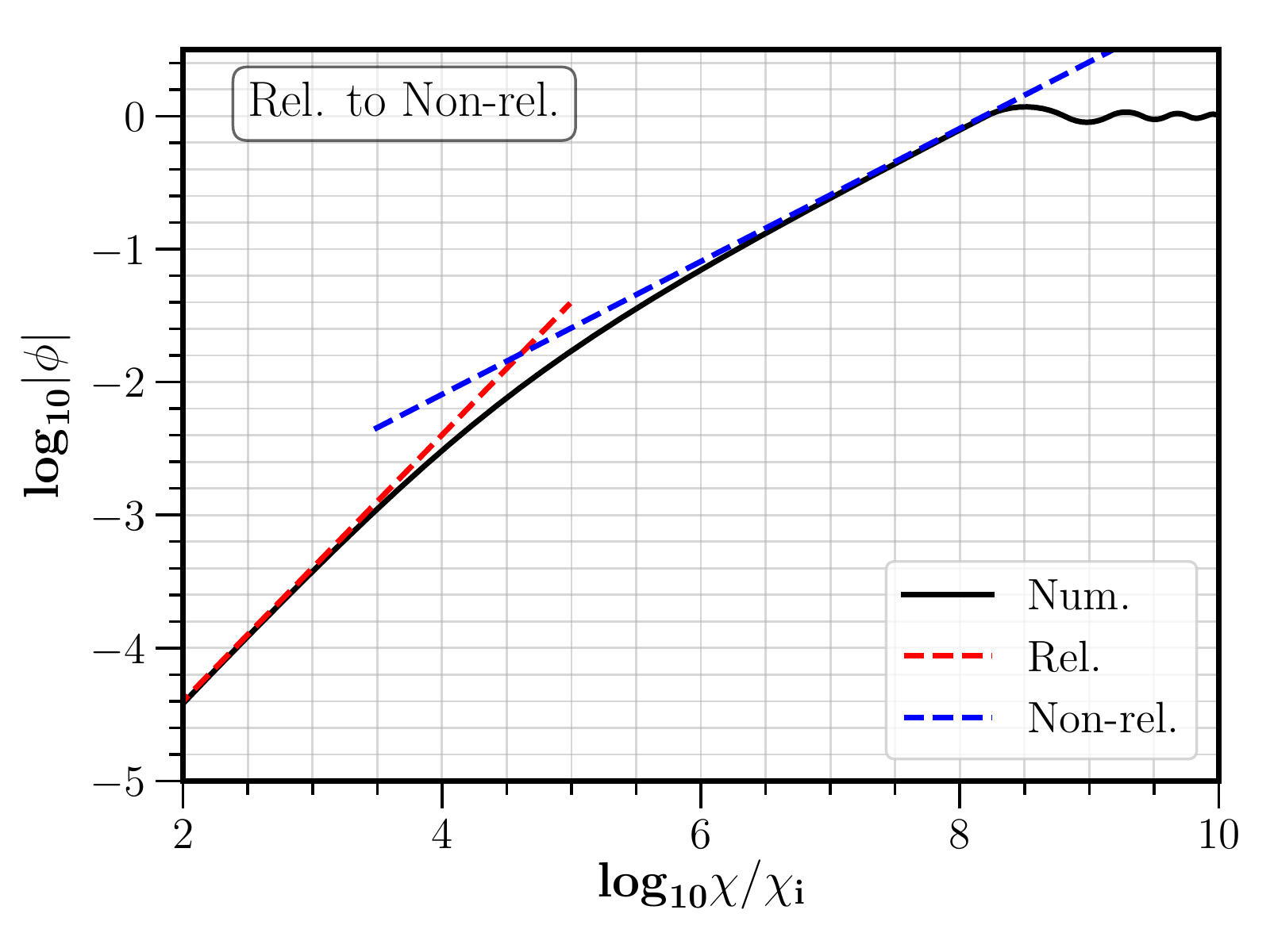}
\caption{Scalar field evolution when the fermion fluid is initially relativistic and becomes non-relativistic before the scalar field saturates. In the figure we chose $\beta=200$, $x_i=10^{-2}$ and $\bar\Omega_{\psi,{\rm r},i}=10^{-7}$. The solid black line is the result of numerical integration while the dashed blue and red lines are the analytical approximations given in the main text. \label{fig:relnonrel}}
\end{figure}

Following from situation \ref{sit:1}, where the fermion is initially relativistic, there are two competing effects. On one hand, since the scalar field is growing, the system is approaching the saturated scalar field regime $m_{\rm eff}=|1+\phi|\sim 0$. On the other hand, the expansion of the universe dilutes the number density of fermions and therefore $x$ from Eq.~\eqref{eq:definitions} increases. Thus, the system may reach the non-relativistic limit where $x\gg1$ before the scalar field saturates. This mixed situation shown in Fig.~\ref{fig:relnonrel}. However, in situation \ref{sit:2} the scalar field saturates before reaching the non-relativistic limit. That is, we consider that $x\ll1$ holds for all times. Requiring that $x\ll1$ at the time when the scalar field saturates roughly yields
\begin{align}\label{eq:conditionrel}
\beta^2\bar\Omega_{\psi,{\rm r},i}\gg1\,.
\end{align}
Assuming that the relativistic limit holds, the Klein-Gordon \eqref{eq:KG2} equation reads
\begin{align}
\frac{d^2\phi}{d\chi^2}+\frac{3}{2\chi}\frac{d\phi}{d\chi}+6x^2_i(1+\phi)\bar\Omega_{\psi,{\rm r},i}\frac{\chi_i}{\chi}=0\,.
\end{align}
The equation above has an exact solution given by
\begin{align}\label{eq:solsit2}
\phi_{\rm (b)}=-1+\frac{\sin \left(
   \omega\sqrt{{\chi}/{\chi_i}}\right)}{
   \omega\sqrt{{\chi}/{\chi_i}}}\,,
\end{align}
where we imposed that $\phi\to 0$ as $\chi\to 0$ and we defined
\begin{align}
\omega\equiv\beta x_i\sqrt{6\bar\Omega_{\psi,{\rm r},i}}\,.
\end{align}
Notice that from Eq.~\eqref{eq:conditionrel} we have that $\omega\lesssim 1$. In Fig.~\ref{fig:rel} we show the numerical solution of \ref{sit:1} and \ref{sit:2} and compare it with the analytical solution \eqref{eq:solsit2}. We find a very good agreement of the numerical results with the analytical approximation. Although we do not show it here explicitly, the density fraction of the fermion \eqref{eq:densityfraction} and of the scalar field are always much smaller than unity and so the radiation component always dominates. Before going into the next subsection, it is worth noting that if the scalar field initially sits at vanishing effective mass, it remains there for the entire evolution. In this case, the fermions are exactly relativistic. 

\subsection{Non-relativistic regime\label{sec:degnonrel}}
Let us turn to the case when the fermions are non-relativistic, that is $m_{\rm eff}\gg N_\psi^{1/3}$. This time, we have that initially $\phi\ll1$ and $x_i\gg1$. The Klein-Gordon equation \eqref{eq:KG2} in this limit reads
\begin{align}
\frac{d^2\phi}{d\chi^2}+\frac{3}{2\chi}\frac{d\phi}{d\chi}+3\bar\Omega_{\psi,{\rm nr},i}\left(\frac{\chi_i}{\chi}\right)^{3/2}=0\,.
\end{align}
The particular solution that yields $\phi\to0$ as $\chi\to 0$ is given by
\begin{align}\label{eq:solsit3}
\phi_{\rm (c)}\approx -\frac{3}{2}\beta^2\bar\Omega_{\psi,{\rm nr},i}\sqrt{\frac{\chi}{\chi_i}}\,.
\end{align}
where the subscript $\rm (c)$ is to emphasize that this solution holds in situation \ref{sit:3}. As in the relativistic case \ref{sit:1}, the scalar field $\phi$ takes negative values and grows due to the attractive force. Note that to make sense of the pivot time $\chi_i$, we require that it is set before the scalar field saturates. This means that in the non-relativistic regime we have that
\begin{align}\label{eq:conditionnr}
\beta^2\bar\Omega_{\psi,{\rm nr},i}\ll 1\,. 
\end{align}
Conditions \eqref{eq:conditionrel} and \eqref{eq:conditionnr} clearly separate the non-relativistic from the relativistic case. Now, since the Fermi gas is initially non-relativistic it stays non-relativistic until the scalar field starts to saturate at $m_{\rm eff}\sim 0$. As it is clear from \eqref{eq:solsit3}, when $\beta>1$ the scalar field always saturates before the non-relativistic fermions dominate the universe. Only if $\beta<1$, the non-relativistic fluid dominates the universe before the saturation. These conclusion holds as long as the scalar field is essentially massless.

\begin{figure}
\centering
\includegraphics[width=0.5\columnwidth]{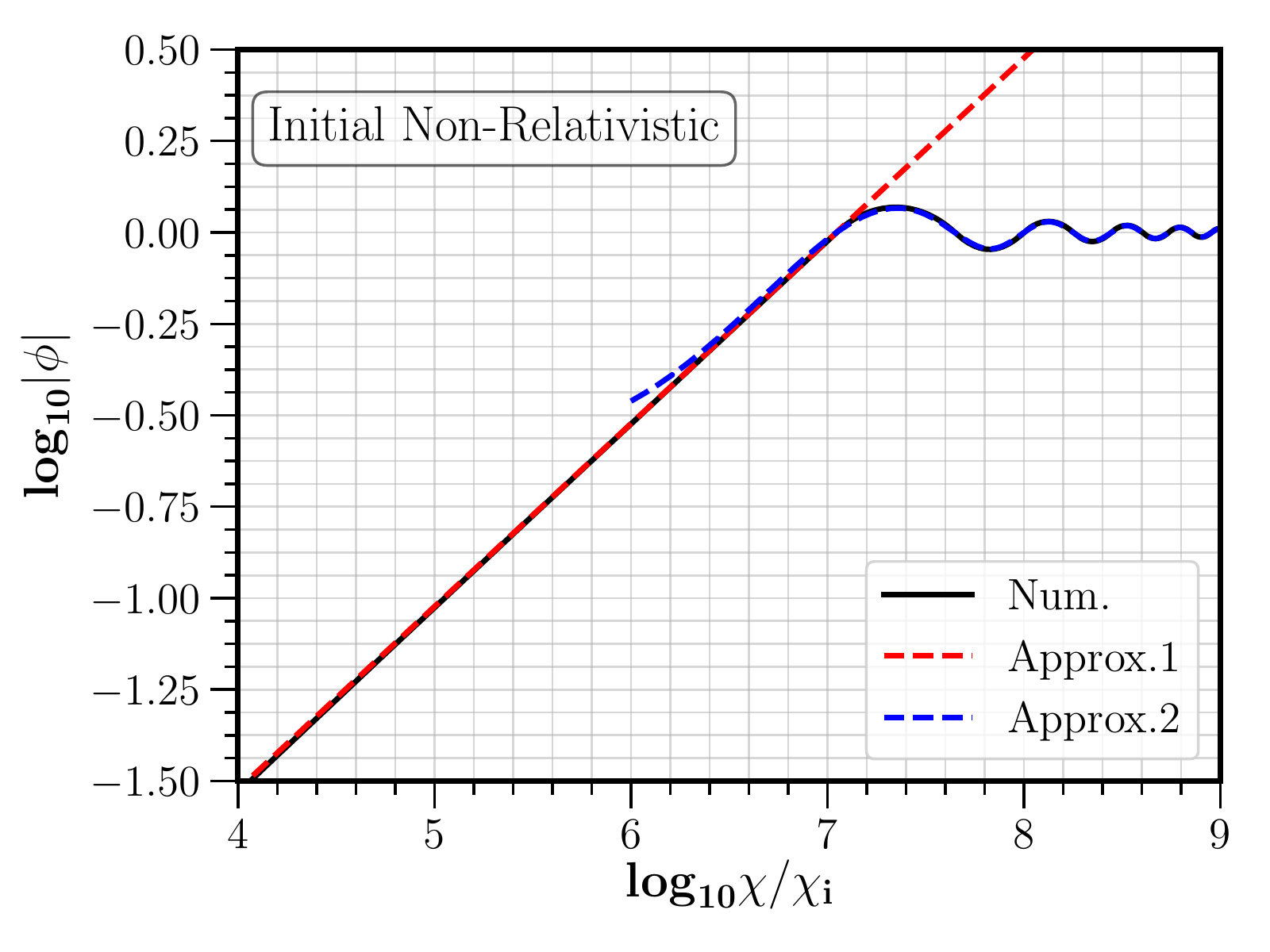}
\caption{Scalar field evolution when the fermion fluid is initially non-relativistic. In the figure we chose $\beta=200$, $x_i=10^{3}$ and $\bar\Omega_{\psi,{\rm nr},i}=5\times10^{-9}$. The solid black line is the result of numerical integration while the dashed blue and red lines are the analytical approximations given in the main text.\label{fig:nonrel}}
\end{figure}

In the exact moment when $m_{\rm eff}\sim0$ the Fermi gas becomes relativistic as $x\sim0$. However, the system does not spend time in the relativistic regime. As we proceed to show, the Fermi gas spends more time in the regime where $x\gg1$ and stays non-relativistic. Interestingly though, due to the time dependence of the effective mass the energy density of the Fermi gas decays as that of a radiation fluid. To convince ourselves that the system does not spend time in the relativistic regime with saturated scalar field, consider the relativistic limit of Eq.~\eqref{eq:KG2} but with non-relativistic initial conditions. Then the Klein-Gordon equation close to $m_{\rm eff}\sim 0$ reads
\begin{align}\label{eq:nonrelreldeg}
\frac{d^2\phi}{d\chi^2}+\frac{3}{2\chi}\frac{d\phi}{d\chi}+\frac{9}{2}x_i(1+\phi)\bar\Omega_{\psi,{\rm nr},i}\frac{\chi_i}{\chi}=0\,.
\end{align}
To solve the above equation we take as initial conditions the first time that $m_{\rm eff}=0$, that is $\phi_{\rm (c)}(\chi_0)=-1$. This condition relates $\chi_0$ to $\chi_i$ by
\begin{align}\label{eq:T0Ti}
\sqrt{\frac{\chi_0}{\chi_i}}=\left(\frac{3}{2}\beta^2\bar\Omega_{\psi,{\rm nr},i}\right)^{-1}\,.
\end{align}
Matching the amplitude and first derivatives of $\phi$ at $\chi_0$, we obtain that the solution close to $m_{\rm eff}\sim 0$ reads
\begin{align}
1+\phi=-\frac{\sin\left[\tilde\omega_{\rm r}\left(\sqrt{\frac{\chi}{\chi_0}}-1\right)\right]}{\tilde\omega_{\rm r}\sqrt{\frac{\chi}{\chi_0}}}
\quad {\rm with}\quad
\tilde\omega_{\rm r}\equiv\frac{1}{\beta}\sqrt{\frac{2x_i}{\bar\Omega_{\psi,{\rm nr},i}}}\,,
\end{align}
where the tilde refers to the initially non-relativistic case and the subscript ``r'' refers to the fact that we are in the relativistic regime when $m_{\rm eff}=0$. We then see that the scalar field oscillates with a frequency proportional to $\tilde\omega_{\rm rel}\gg1$, where we used that $x_i\gg 1$ and $\beta^2\bar\Omega_{\psi,{\rm nr},i}<1$ \eqref{eq:conditionnr}. Although the amplitude of the oscillations is small, we find that the system exits the relativistic regime very fast. First, because the frequency is large and, second, because the amplitude of the oscillations yield a value of $x$ which quickly exceeds unity, as it is given by
\begin{align}
x\approx\frac{x_i}{\tilde\omega_{\rm r}}\left|\sin\left[\tilde\omega_{\rm r}\left(\sqrt{\frac{\chi}{\chi_0}}-1\right)\right]\right|\approx x_i\left(\sqrt{\frac{\chi}{\chi_0}}-1\right)\,,
\end{align}
where in the last step we expanded for small argument. This is also clear if we look at the rate of change of the parameter $x$ per Hubble time, explicitly given by
\begin{align}
\frac{dx}{d\ln a}\bigg|_{\chi=\chi_0}=2\chi\frac{dx}{d\chi}\bigg|_{\chi=\chi_0}={x_i}\gg1\,.
\end{align}
We find a large variation of $x$ per Hubble time since $x_i\gg 1$.
Thus, we shall use the non-relativistic approximation instead to derive analytical formulas. We confirm these expectations numerically and are shown in Fig.~\ref{fig:nonrel}.

\begin{figure}
\centering
\includegraphics[width=0.5\columnwidth]{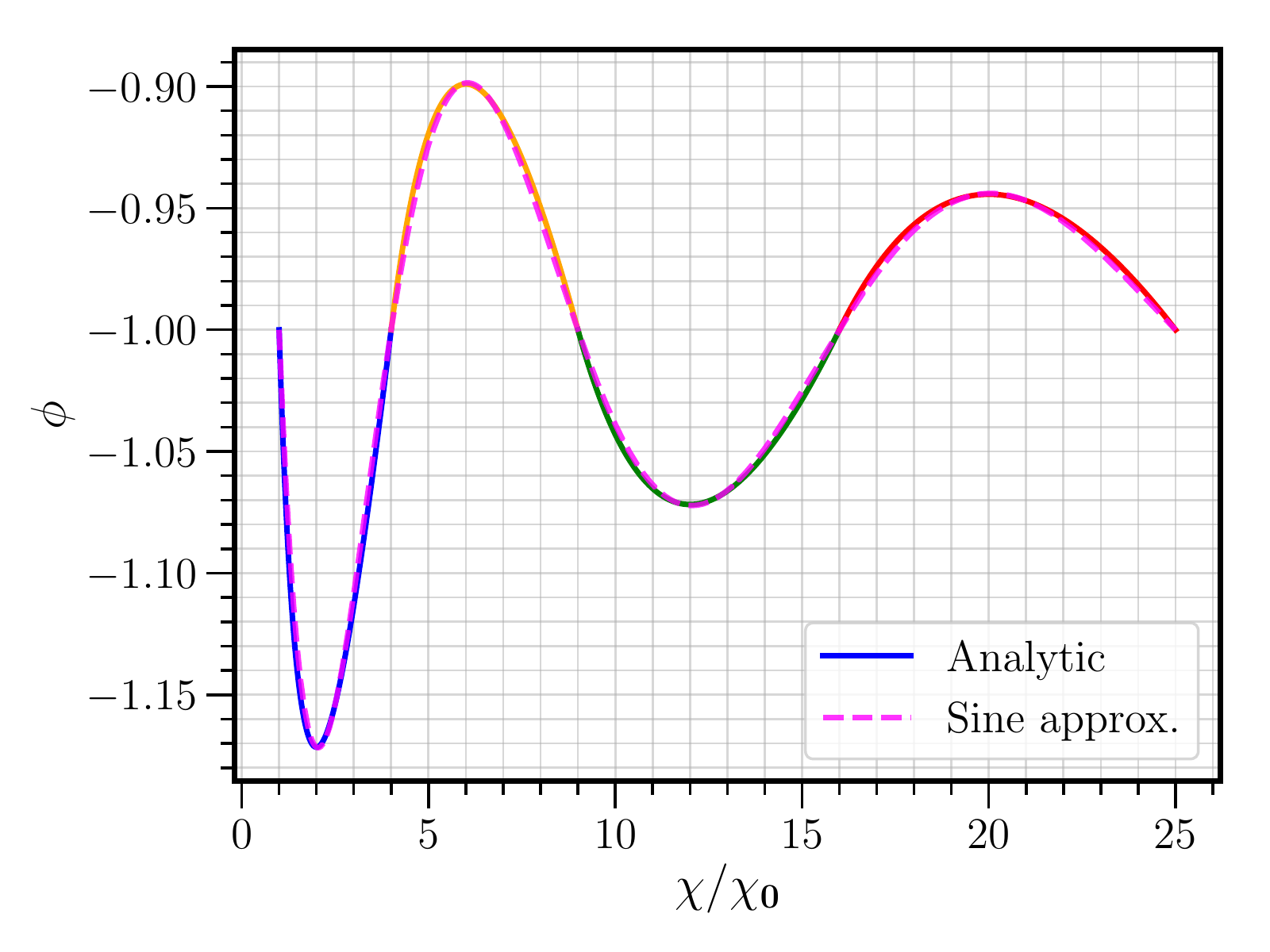}
\caption{Scalar field solution for the case \ref{sit:4} when the fermion is non-relativistic and the scalar field is saturated. $\chi_0$ corresponds to the first time when the scalar field reaches unity, i.e. $\phi(\chi_0)=-1$. In solid lines we show the analytical solution by assuming that the fermion gas stays non-relativistic even though there are instants of a relativistic equation of state parameter. Since the analytical solution in solid lines is unpractical due to sign change of the source term, we present a sinusoidal approximation that works very well shown in dashed lines. The change in the color of the solid lines is to illustrate that the sign of the source changes. \label{fig:approx}}
\end{figure}

In the non-relativistic regime we have that in general the Klein-Gordon equation \eqref{eq:KG2} reads
\begin{align}\label{eq:KGNONREL}
\frac{d^2\phi}{d\chi^2}+\frac{3}{2\chi}\frac{d\phi}{d\chi}+3\frac{|1+\phi|}{1+\phi}\bar\Omega_{\psi,{\rm nr},i}\left(\frac{\chi_i}{\chi}\right)^{3/2}=0\,.
\end{align}
We see that $\phi$ must be an oscillating function due to the change of sign in the source term everytime that $\phi$ crosses $-1$. So even if the Fermi gas stays non-relativistic, the scalar field oscillates around $m_{\rm eff}\sim 0$. Eq.~\eqref{eq:KGNONREL} may be solved recursively given initial conditions.
In terms of $\chi_0$ the solution of Eq.~\eqref{eq:KGNONREL} at the $j$-th time that $\phi(\chi_j)=-1$, with $j>0$, is given by
\begin{align}\label{eq:solsit42}
1+\phi_j=(-1)^{j+1} \sqrt{\frac{\chi}{\chi_0}}+ C_j+\tilde C_j\sqrt{\frac{\chi_0}{\chi}}\,.
\end{align}
Solving $\phi(\chi_j)=-1$ we find the value of $\chi_j$ where $m_{\rm eff}=0$, explicitly given by
\begin{align}
\sqrt{\frac{\chi_j}{\chi_0}}=\frac{1}{2}\left((-1)^{j}C_j+\sqrt{C_j^2+(-1)^{j}4\tilde C_j}\right)\,.
\end{align}
Then the matching conditions at $\chi_j$ yield a recursive relation among the coefficients, that reads
\begin{align}
 C_{j+1}&=C_j+(-1)^{j+1}4\frac{\chi_j}{\chi_0}\,,\\
 \tilde C_{j+1}&=\tilde C_j+(-1)^{j}2\sqrt{\frac{\chi_j}{\chi_0}}\,.
\end{align}
Such recursive relation start at $j=1$ where
\begin{align}
C_1=-3\,,\quad \tilde C_1=2\,.
\end{align}
In Fig.~\ref{fig:approx} we show in solid lines the analytical solution for two oscillations. The change in color stands for change of sign of $1+\phi$. A useful relation to solve the recurrence relations is the following:
\begin{align}
\sqrt{\frac{\chi_{j+1}}{\chi_0}}=3\sqrt{\frac{\chi_{j}}{\chi_0}}+(-1)^{j+1}C_{j}\,.
\end{align}
Then, we find that the recurrence relations are solved in terms of $j$ as
\begin{align}
\sqrt{\frac{\chi_{j}}{\chi_0}}=(1+j)\,,\quad C_j=(-1)^j (2 j + 1)\quad{\rm and}\quad \tilde C_j=(-1)^{1 + j} j (1 + j)\,.
\end{align}

\begin{figure}
\centering
\includegraphics[width=0.5\columnwidth]{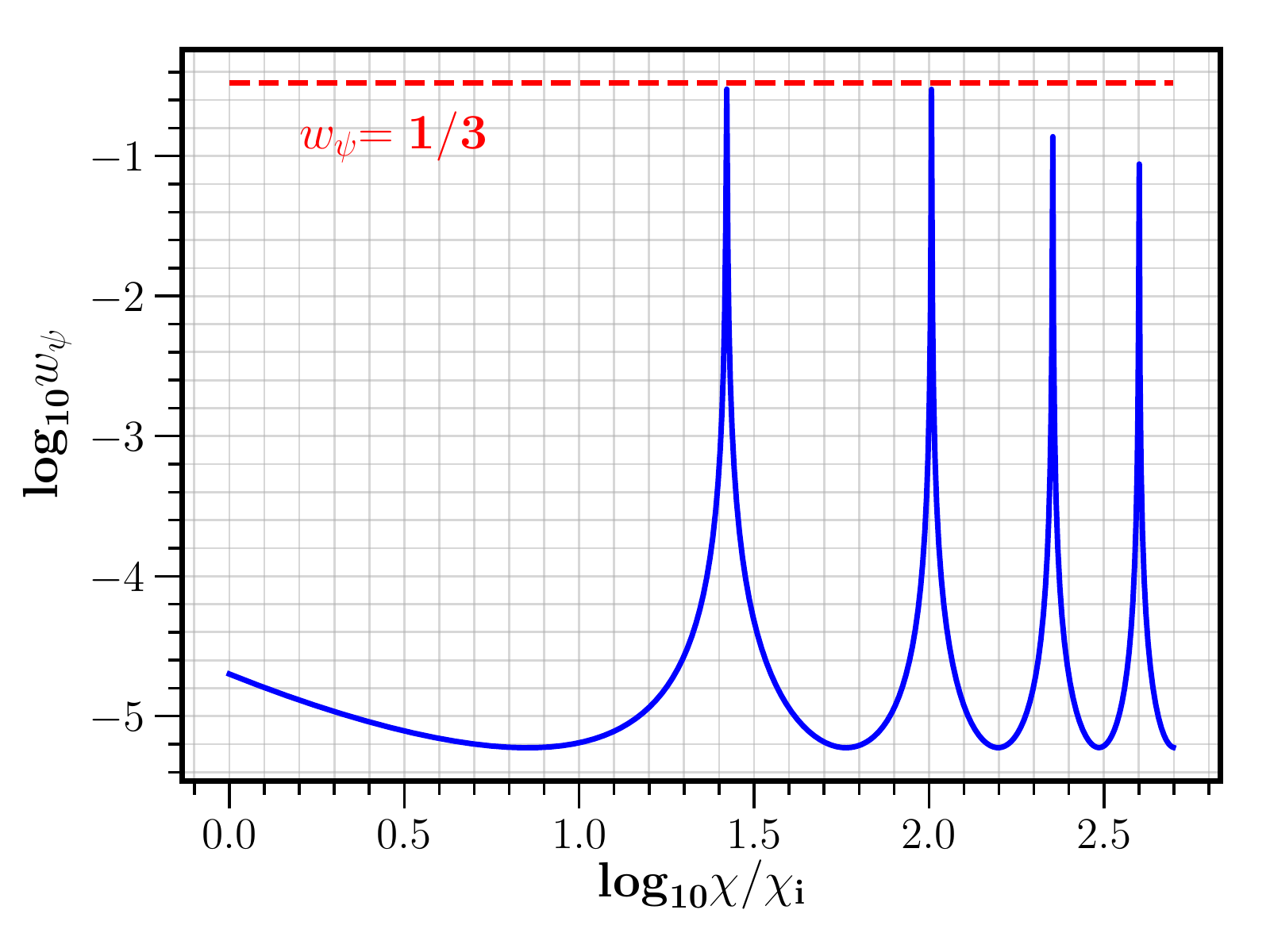}
\caption{Equation of state parameter of the fermion fluid in the initially non-relativistic case. In the figure we chose $\beta=200$, $x_i=10^{2}$ and $\bar\Omega_{\psi,{\rm nonrel},i}=5\times10^{-6}$. See how the fermion becomes relativistic for very short periods of time and, hence, the presence of spikes. However, the fermion fluid is non-relativistic for practical purposes. The last peaks seem not to reach $w_\psi=1/3$ due to a poor numerical resolution around the spikes.\label{fig:w}}
\end{figure}

For practical purposes though, it proves more useful to find an approximate description of the analytical solution. First, from the solution \eqref{eq:solsit4}, we know that the scalar field amplitude grows with $\sqrt{\chi}$ and decays with $1/\sqrt{\chi}$. Second, we note that the period of the oscillation follows $\sqrt{\chi/\chi_0}$ since $\sqrt{\chi_j/\chi_0}$ increase by $1$ every half oscillation. Thus, the analytical solution resembles very much a sine function, since we have that initially $1+\phi(\chi_0)=0$. We give the following formula as an approximated solution:
\begin{align}\label{eq:solsit4bis}
1+\phi_{(d)}=\frac{4-3\sqrt{2}}{\sin(\sqrt{2}\pi)}\frac{\sin\left(\pi\sqrt{{\chi}/{\chi_0}}\right)}{\sqrt{{\chi}/{\chi_0}}}\,,
\end{align}
where the amplitude has been fixed by matching the maximum at $T=2T_0$. This is then translated to the pivot time as
\begin{align}\label{eq:solsit4}
1+\phi_{(d)}=\pi\frac{4-3\sqrt{2}}{\sin(\sqrt{2}\pi)}\frac{\sin\left(\tilde\omega_{\rm nr}\sqrt{{\chi}/{\chi_i}}\right)}{\tilde\omega_{\rm nr}\sqrt{{\chi}/{\chi_i}}}
\quad{\rm with}\quad
\tilde\omega_{\rm nr}=\frac{3\pi}{2}\beta^2\bar\Omega_{\psi,{\rm nr},i}.
\end{align}
We compare the exact analytical formula with the approximation in Fig.~\ref{fig:approx}. We find an extremely good agreement between the two and that holds for the time run in the numerical integration of the exact solution as in Fig.~\ref{fig:nonrel}.

Once we have an analytical solution to scalar field, we shall discuss the behavior of the Fermi fluid. We numerically confirm in Fig.~\ref{fig:w} that the fermi gas spends more time in the non-relativistic regime than in the relativistic regime. This is also clear by computing the ratio $x$, that is
\begin{align}
x(\chi>\chi_0)=\frac{m_{\rm eff}}{N_\psi^{1/3}}=x_i |1+\phi|\sqrt{\frac{\chi}{\chi_i}}=x_i\frac{\pi}{{\tilde\omega_{\rm nr}}}\frac{4-3\sqrt{2}}{\sin(\sqrt{2}\pi)}\left|{\sin\left(\tilde\omega_{\rm nr}\sqrt{{\chi}/{\chi_i}}\right)}\right|\,.
\end{align}
Therefore, the oscillation average is
\begin{align}
\langle x\rangle(\chi>\chi_0) =\frac{2x_i}{{\tilde\omega_{\rm nr}}}\frac{4-3\sqrt{2}}{\sin(\sqrt{2}\pi)}\approx\frac{x_i}{{2\tilde\omega_{\rm nr}}}\gg 1\,,
\end{align}
where we used Eq.~\eqref{eq:T0Ti}. This implies that the Fermi gas is on average well described by the non-relativistic limit. This is also clear from the evolution of the equation of state parameter shown in Fig.~\ref{fig:w}. We see that $w_\psi$ is very small except for short periods of time where it reaches $w_\psi=1/3$. Similarly for the energy density of the fermi gas we find
\begin{align}
\rho_{\psi}(\chi>\chi_0)\approx \frac{1}{3\pi^2}{m_{\psi} N_\psi}|1+\phi|=\rho_{\psi,{\rm nr},i}\frac{\pi}{{\tilde\omega_{\rm nr}}}\frac{4-3\sqrt{2}}{\sin(\sqrt{2}\pi)}\left|{\sin\left(\tilde\omega_{\rm nr}\sqrt{{\chi}/{\chi_i}}\right)}\right|\left(\frac{\chi_i}{\chi}\right)^{2}\,.
\end{align}
Taking the oscillation average we find that
\begin{align}
\langle \Omega_\psi\rangle(\chi>\chi_0)\approx\frac{2}{{\tilde\omega_{\rm nr}}}\frac{4-3\sqrt{2}}{\sin(\sqrt{2}\pi)}\bar \Omega_{\psi,{\rm nr},i}\approx\frac{1}{{2\tilde\omega_{\rm nr}}}\bar \Omega_{\psi,{\rm nr},i}=\frac{1}{3\pi\beta^2}\,
\end{align}
where we used again Eq.~\eqref{eq:T0Ti}. We see that while the Fermi gas is in the non-relativistic limit, the energy density decays as that of the radiation fluid due to the decreasing effective mass. In this regime we also have that
\begin{align}
\Omega_\varphi\approx\frac{\pi^2}{3\beta^2}\left(\frac{4-3\sqrt{2}}{\sin(\sqrt{2}\pi)}\right)^2\cos^2\left(\tilde\omega\sqrt{{\chi}/{\chi_i}}\right)\,,
\end{align}
and therefore the average is given by
\begin{align}
\langle \Omega_\varphi\rangle(\chi>\chi_0)\approx\frac{\pi^2}{6\beta^2}\left(\frac{4-3\sqrt{2}}{\sin(\sqrt{2}\pi)}\right)^2\approx\frac{\pi^2}{96\beta^2}\,.
\end{align}

\subsection{Massive scalar field regime \label{subsec:mass}}

Here we briefly explore the possibility that the massive fermions are the dominant component of dark matter. Thus, we shall focus solely on the non-relativistic case studied in Sec.~\ref{sec:degnonrel} after the scalar field saturated. As a rough approximation we assume that the scalar field is exactly massive right after the time $t_*$ when $H_*=M$. Let us also assume that $t_*\gg t_0$ so that the scalar field spends enough time in the saturation regime. In this case, a good approximation is that $\phi\approx -1$ or, equivalently, $\varphi\approx-M_{\rm pl}\beta^{-1}$. With this initial condition at $t=t_*$, the solution of Eq.~\eqref{eq:KLEIN} in the massive scalar field regime and in a radiation dominated universe is given by
\begin{align}\label{eq:varphimassive}
\varphi\approx -\frac{M_{\rm pl}}{\beta}\left(\frac{t}{t_*}\right)^{-3/4}\cos\left(M(t-t_*)\right)\,.
\end{align}
Furthermore, we can also assume that the Yukawa interaction is shut off when the scalar field becomes massive enough at $H_*=M$. This occurs fairly quickly as the Yukawa interaction becomes a short range interaction. This means that the scalar field decays and the density fraction of the non-relativistic fluid grows. Using Eq.~\eqref{eq:varphimassive} the energy density of the scalar field from the coherent oscillations is given by
\begin{align}
\rho_\varphi\approx \frac{M^2M_{\rm pl}^2}{2\beta^2}\left(\frac{t}{t_*}\right)^{-3/2}\,.
\end{align}
In this way we find that the density fraction of the scalar field and fermions are respectively given by
\begin{align}
\Omega_\varphi\approx \frac{1}{6\beta^2}\left(\frac{t}{t_*}\right)^{1/2}\,,\quad\Omega_\psi\approx \frac{1}{3\pi\beta^2}\left(\frac{t}{t_*}\right)^{1/2}\,.
\end{align}
Their ratio gives $\Omega_\varphi\approx \frac{2}{\pi}\Omega_\psi$. This means that, in this rough approximation, the scalar field makes up to $36\%$ of the total dark matter. 

With these results, let us derive some constraints on the model parameters by requiring that this model explains the CDM abundance. To do that, we compute the abundance of CDM at $H_*=M$ leading us to
\begin{align}
\Omega_{\rm CDM,*}\approx 3\times 10^{-16}\left(\frac{M}{10^{-6}\,{\rm GeV}}\right)^{-1/2}\left(\frac{g(T_*)}{106.75}\right)^{1/4}\left(\frac{g_s(T_*)}{106.75}\right)^{-1/3}\,,
\end{align}
where we used that $g(T_*)\approx 106.75$ and
\begin{align}
T_*=8.43\times 10^{5}\,{\rm GeV}\left(\frac{M}{10^{-6} {\rm GeV}}\right)^{1/2}\,.
\end{align}
We find that in order to have right abundance of dark matter, composed of the sum of the fermions and the scalar field, we need
\begin{align}\label{eq:Mandm}
M\approx43\,{\rm GeV}\left(\frac{y}{10^{-2}}\right)^4\left(\frac{m_\psi}{10^7\,\rm GeV}\right)^{-4}\,.
\end{align}
Thus, this mechanism for generating the right abundance dark matter implies that a large fraction of dark matter is in the form of a scalar field.

\section{Non-degenerate Fermi gas\label{sec:nondeg}}

In the previous section we have found that, in the degenerate limit, there is a scaling solution after the scalar field saturates. However, in a more realistic situation the Fermi gas might be in equilibrium with the dark radiation bath. Thus, in this section we consider a Fermi gas at finite temperature $T$ and chemical potential $\mu$. Let us recall that in the non-degenerate case, the chemical potential $\mu$ is not completely fixed by the Fermi energy and one has to find its time dependence by the number and energy density conservations, together with the time dependence of $T$. Furthermore, to be more precise, we should be careful about what are the interactions and how the equilibrium condition is met, specially in the transition from the relativistic to non-relativistic regime. Such numerical calculation is out of the scope of this paper and, thus, we shall only focus on the relativistic and non-relativistic limits separately and analytically. Here, we show that if the Fermi gas is in equilibrium, the scaling solution also holds for the non-degenerate case. We may anticipate that this is the case in the non-relativistic limit as the temperature decays and the Fermi gas essentially becomes degenerate.

As in Sec.~\ref{sec:deg} we shall focus on the relativistic regime first and then turn to the non-relativistic case. Contrary to the degenerate case, the limit of relativistic and non-relativistic is determined by the ratio of the effective mass and the temperature, i.e. $m_{\rm eff}/T$. We then expand the integrals \eqref{eq:thermo}-\eqref{eq:p} for either $m_{\rm eff}\ll T$ (relativistic) or $m_{\rm eff}\gg T$ (non-relativistic). For the moment, we do not assume any hierarchy between $\mu$ and $T$. We only require that $\mu\geq m_{\rm eff}$. Details of the expansion and improved analytical approximations are given in Appendix~\ref{app:series}.

\subsection{Relativistic regime\label{sec:nondegrel}}

In the relativistic regime, the number and energy densities up to next to leading order in $m_{\rm eff}/T\ll1$ are given by
\begin{align}
n_{\psi,\rm r}&=\frac{T^3}{3\pi^2}\left(\pi^2\frac{\mu}{T}+\frac{\mu^3}{T^3}-\frac{3}{2}\frac{m_{\rm eff}^2}{T^2}\frac{\mu}{T}\right)+O\left(\frac{m_{\rm eff}^4}{T^4}\right)\,,\\
\rho_{\psi,\rm r}&=\frac{7}{4}\frac{\pi^2}{15}T^4\left(1+\frac{30}{7\pi^2}\frac{\mu^2}{T^2}+\frac{15}{7\pi^4}\frac{\mu^4}{T^4}-\frac{5m_{\rm eff}^2}{7\pi^4T^2}\left(\pi^2+3\frac{\mu^2}{T^2}\right)\right)+O\left(\frac{m_{\rm eff}^4}{T^4}\right)\,.
\end{align}
We also find that the source term to the Klein-Gordon equation is proportional to
\begin{align}
\rho_{\psi,\rm r}-3P_{\psi,\rm r}\approx \frac{m_{\rm eff}^2T^2}{6\pi^2}\left(\pi^2+3\frac{\mu^2}{T^2}\right)+O\left(\frac{m_{\rm eff}^4}{T^4}\right)\,.
\end{align}
We see that the pressure at leading order is $P_\psi=\rho_\psi/3$, as usual. From the leading order terms in the number and energy density conservation we conclude that
\begin{align}
T_{\rm r}= T_i\frac{a_i}{a}+O\left(\frac{m_{\rm eff}^2}{T^2}\right)\,,\quad\mu_{\rm r}=\mu_i\frac{a_i}{a}+O\left(\frac{m_{\rm eff}^2}{T^2}\right)\,,
\end{align}
where we stopped at leading order since it is enough to solve the Klein-Gordon equation. The ratio $\mu_i/T_i$ is then determined once $T_i$ and $n_{\psi,i}$ are given. We shall assume that $\mu_i/T_i\ll1$ as usually one expects $\mu\propto m_{\rm eff}$.  This means that a leading order we have
\begin{align}
\frac{\mu_i}{T_i}\approx\frac{3n_{\psi,i}}{T_i^3}\,.
\end{align}

With these results we can solve the Klein-Gordon equation \eqref{eq:KLEIN}, which in the massless regime and using the previous redefinitions \eqref{eq:definitions} reads
\begin{align}\label{eq:KGnondegrel}
\frac{d^2\phi}{d\chi^2}+\frac{3}{2\chi}\frac{d\phi}{d\chi}+\left(1+\phi\right)\frac{30}{7\pi^2}\frac{m_\psi^2}{T_i^2}\Omega_{\psi,\rm r, i}\frac{\chi_i}{\chi}=0\,.
\end{align}
Eq.~\eqref{eq:KGnondegrel} has the same qualitative behavior as the degenerate relativistic case of Sec.~\ref{sec:degrel}. In this case, the general solution reads
\begin{align}
1+\phi_{\rm r}=\frac{\sin\left(\lambda\sqrt{\chi/\chi_i}\right)}{\lambda\sqrt{\chi/\chi_i}}
\quad{\rm with}\quad
\lambda\equiv\frac{m_\psi}{T_i}\frac{\beta}{\pi}\sqrt{\frac{30}{7}\Omega_{\psi,\rm r, i}}\,.
\end{align}
For $\chi\ll\chi_i$ the scalar field grows as
\begin{align}
\phi_{\rm r}\approx -\frac{\lambda^2}{6}\frac{\chi}{\chi_i}\,.
\end{align}
This means that the scalar field reaches unity roughly at
\begin{align}
\frac{\chi_0}{\chi_i}\approx\frac{6}{\lambda^2}\,.
\end{align}
As in the degenerate case, depending on the parameters, the Fermi gas might enter the non-relativistic regime before saturating the scalar field. To check that, we evaluate the ratio $m_{\psi}/T_{\rm r}$ roughly when the scalar field saturates, which yields
\begin{align}
\frac{m_{\psi}}{T_{\rm r}}\Big|_{\chi_0}\approx\frac{\pi}{\beta}\sqrt{\frac{7}{5\Omega_{\psi,\rm r, i}}}\,.
\end{align}
Thus, the system stays in the relativistic regime for the whole evolution if ${m_{\psi}}/{T_{\rm r}}|_{\chi_0}\ll1$ which requires
\begin{align}\label{eq:conditionrel2}
\beta^2\Omega_{\psi,\rm r,i}\gg1\,.
\end{align}
This is the same requirement than in the relativistic degenerate case \eqref{eq:conditionrel}. Thus, if we start with initial conditions that satisfy \eqref{eq:conditionrel2}, the Fermi gas stays indefinitely in the relativistic regime and the scalar field oscillates around $m_{\rm eff}\sim 0$.

\subsection{Non-relativistic regime \label{sec:nondegnonrel}}

Let us study in more detail the non-relativistic regime, that is when $m_{\rm eff}\gg T$. For practical purposes we start by presenting the form of the pressure of the non-relativistic Fermi gas, which reads
\begin{align}
P_{\psi,\rm nr}\approx- T\frac{\left(m_{\rm eff}T\right)^{3/2}}{\sqrt{2}\pi^{3/2}}{\rm Li}_{5/2}\left(-e^{-(m_{\rm eff}-\mu)/T}\right)\,,
\end{align}
where ${\rm Li}_s(x)$ is the polylogarithm function of order $s$ (see Appendix~\ref{app:series} for its asymptotic limits). For the number and energy density we respectively obtain
\begin{align}
n_{\psi,\rm nr}&\approx- \frac{\left(m_{\rm eff}T\right)^{3/2}}{\sqrt{2}\pi^{3/2}}{\rm Li}_{3/2}\left(-e^{-(m_{\rm eff}-\mu)/T}\right)+\frac{15}{8m_{\rm eff}}{P_{\psi,\rm nr}}\,,\\
\rho_{\psi,\rm nr}&\approx- m_{\rm eff}\frac{\left(m_{\rm eff}T\right)^{3/2}}{\sqrt{2}\pi^{3/2}}{\rm Li}_{3/2}\left(-e^{-(m_{\rm eff}-\mu)/T}\right)+\frac{3}{2}P_{\psi,\rm nr}=m_{\rm eff} n_{\psi,\rm nr}+\frac{3}{2}P_{\psi,\rm nr}\label{eq:rhononrelnondeg}\,,
\end{align}
where we see that the next to leading order terms are proportional to the pressure. It should be noted that in the above expressions we neglected exponentially suppressed terms proportional to $e^{-(m_{\rm eff}+\mu)/T}$. As usual in the non-relativistic case we see that the pressure is negligible and that $\rho_{\psi,{\rm nr}}\approx m_{\rm eff} n_{\psi,{\rm nr}}$. From the latter we see that, at leading order, the energy conservation yields the number density conservation. Looking at the number density conservation for general values of $\mu$ we conclude that
\begin{align}\label{eq:temprel}
T_{\rm nr}\approx T_i\left(\frac{a_i}{a}\right)^{2}\left(\frac{m_{{\rm eff},i}}{m_{\rm eff}}\right)\,,\quad\mu_{\rm nr}\approx m_{\rm eff}+\mu_i\left(\frac{a_i}{a}\right)^{2}\left(\frac{m_{{\rm eff},i}}{m_{\rm eff}}\right)\,.
\end{align}

We are ready to study the Klein-Gordon equation in the non-relativistic limit. Using Eq.~\eqref{eq:rhononrelnondeg} we find that the scalar field equation has the same qualitative dynamics as in the degenerate case \eqref{eq:KGNONREL}, that is
\begin{align}\label{eq:KGnondegnonrel}
\frac{d^2\phi}{d\chi^2}+\frac{3}{2\chi}\frac{d\phi}{d\chi}+3\frac{|1+\phi|}{1+\phi}\Omega_{\psi,{\rm nonrel},i}\left(\frac{\chi_i}{\chi}\right)^{3/2}=0\,.
\end{align}
Analogous with the degenerate case of Sec.~\ref{sec:degnonrel}, we know that the scalar field grows according to Eq.~\eqref{eq:solsit3} until it reaches $m_{\rm eff}\sim 0$. The main difference, however, is that the relativistic and non-relativistic regimes are separated by a different parameter. In the degenerate case we have $m_{\rm eff}/n_\psi^{1/3}$ and in the non-degenerate case it is $m_{\rm eff}/T$. Thus, the question is whether in the non-degenerate case, the system also remains non-relativistic. 

To check whether the Fermi gas stays non-relativistic, let us use the results of Sec.~\ref{sec:degnonrel} where we found that initially the field grows according to Eq.~\eqref{eq:solsit3} and saturates roughly at $\chi=\chi_0$ given by Eq.~\eqref{eq:T0Ti}. Let us also assume for simplicity that initially the chemical potential is very close to the effective mass, that is we have $\mu_i/T_i\ll1$ in Eq.~\eqref{eq:temprel}. This implies that the initial number and energy densities are well approximated by
\begin{align}
n_{\psi,\rm nr,i}\approx \frac{\left(m_{\psi}T_i\right)^{3/2}}{\sqrt{2}\pi^{3/2}}e^{\mu_i/T_i}\,,\quad
\rho_{\psi,\rm nr,i}\approx m_{\psi}n_{\psi,\rm nr,i}\,.
\end{align}
Instead, when the effective mass is very close to $m_{\rm eff}\sim 0$ and the scalar field saturates around $\phi\sim -1$ at $\chi=\chi_0$, the Klein-Gordon equation is well approximated by the relativistic limit \eqref{eq:KGnondegrel} evaluated at $\chi\sim\chi_0$. Keeping in mind that the temperature in the non-relativistic regime goes as $a^{-2}\propto \chi^{-1}$ \eqref{eq:temprel}, we find that close to $m_{\rm eff}\sim 0$ the scalar field equation in terms of the initial non-relativistic energy density is given by
\begin{align}\label{eq:KGnondegnonrelrel}
\frac{d^2\phi}{d\chi^2}+\frac{3}{2\chi}\frac{d\phi}{d\chi}+({1+\phi})e^{-\mu_i/T_i}\pi^{3/2}\sqrt{\frac{T_i}{2m_\psi}}\Omega_{\psi,{\rm nr},i}\left(\frac{\chi_i}{\chi_0}\right)^2\frac{\chi_0}{\chi}=0\,.
\end{align}
This equation is qualitatively the same than the corresponding one in the degenerate case \eqref{eq:nonrelreldeg} except for the suppression factor $T/m_\psi$ and for the fact that $T(\chi<\chi_0)$ decays faster than $a^{-1}$. The solution to \eqref{eq:KGnondegnonrelrel} with matching conditions at $\chi=\chi_0$ is given by
\begin{align}
1+\phi=-\frac{\sin\left[\tilde\lambda_{\rm r}\left(\sqrt{\frac{\chi}{\chi_0}}-1\right)\right]}{\tilde\lambda_{\rm r}\sqrt{\frac{\chi}{\chi_0}}}\quad {\rm with}\quad
\tilde\lambda_{\rm r}\equiv \frac{\pi^{3/4}}{2^{1/4}}e^{-\tfrac{\mu_i}{2 T_i}}{\beta\sqrt{\Omega_{\psi,{\rm nr},i}}}\left(\frac{T_i}{m_\psi}\right)^{1/4}\,.
\end{align}
This time, contrary to the degenerated case, the frequency given by $\tilde\lambda_{\rm r}$ is small. Nevertheless, a closer inspection of the perturbative expansion factor ${m_{\rm eff}}/{T}$ at times $\chi>\chi_0$ roughly yields
\begin{align}
\frac{m_{\rm eff}}{T}\Big|_{\chi>\chi_0}\approx\frac{m_{\psi}}{\tilde\lambda_{\rm r} T_i}\left|\sin\left[\tilde\lambda_{\rm r}\left(\sqrt{\frac{\chi}{\chi_0}}-1\right)\right]\right|\approx \frac{m_\psi}{T_i}\left(\sqrt{\frac{\chi}{\chi_0}}-1\right)\,,
\end{align}
where in the last step we expanded for small argument. We see that because initially $m_\psi/T_i\gg1$ we have that the amplitude of ${m_{\rm eff}}/{T}|_{\chi>\chi_0}\gg1$. Alternatively, if we look at the time variation we have that
\begin{align}
\frac{d}{d\ln a}\frac{m_{\rm eff}}{T}\bigg|_{\chi=\chi_0}\approx\frac{m_\psi}{ T_i}\gg1\,.
\end{align}
Thus, although $\tilde\lambda_{\rm r}$ is very small, the time dependence of ${m_{\rm eff}}/{T}$ is very strong. We conclude that the Fermi gas leaves the relativistic regime very quickly as in the degenerate non-relativistic case. Thus, we can safely use the approximation given by Eq.~\eqref{eq:solsit4bis} and the results that followed in Secs.~\ref{sec:degnonrel} and \ref{subsec:mass}. To summarize, once the system enters the scalar field saturated regime, the energy density of the fermions decays as radiation but the non-relativistic approximation is still very good. In the non-degenerate case, the non-relativistic approximation should be even better than in the degenerate case as $m_{\rm eff}/T\propto \sqrt{{\chi}/{\chi_0}}$ and keeps growing. 

\section{Conclusions\label{sec:conclusions}}

Long range scalar forces in the early universe may be responsible for the formation of PBHs \cite{Amendola:2017xhl,Flores:2020drq}, or very compact objects \cite{Savastano:2019zpr}. As a first step towards more precise predictions, we studied the general relativistic formulation of a Fermi gas with a scalar field dependent mass in the early universe. By thermodynamical arguments, we found that the consistent coupling term between the fermions and the scalar field, at level of the equations of motion, is given by $\left({\partial P_\psi}/{\partial\varphi}\right)_{T,\mu}$ in the grand canonical ensemble. The grand canonical ensemble is the most suitable description of the Fermi gas in cosmological situations where thermal and chemical equilibrium is assumed. Nevertheless, if one insists in working in the microcanonical ensemble, then the correct coupling is given by $\left({\partial \rho_\psi}/{\partial\varphi}\right)_{s_\psi,n_\psi}$. In that case, however, one must be careful to take the variation at fixed entropy $s_\psi$ and net number density $n_\psi$. We then found that $\left({\partial P_\psi}/{\partial\varphi}\right)_{T,\mu}$ is proportional to the trace of the energy momentum tensor corresponding the Fermi gas perfect fluid, which is consistent with earlier calculations without the statistical physics description \cite{Amendola:1999er,Amendola:2017xhl}.

In this paper, we then focused on the coupled background dynamics of the fermions and the scalar field in a radiation dominated universe. We considered a Yukawa type interaction but our methodology can be easily extended to general couplings. We first gained intuition in Sec.~\ref{sec:deg} with the analytically solvable degenerate gas case, that is the $T\to 0$ limit. We found that for an initially vanishing and massless scalar field, the scalar field grows towards minimizing the effective mass and saturates when the effective mass vanishes. The scalar field then oscillates around the saturation point, i.e. at around $m_{\rm eff}=0$. We provided analytical solutions for the two limiting cases of interest, that is the relativistic and non-relativistic regimes. 

On one hand, the most interesting result is that in the case of an initially non-relativistic Fermi gas, the oscillations of the scalar field around $m_{\rm eff}\sim0$ (which is the exact relativistic point) still yield an effectively non-relativistic Fermi gas. As we have shown, both numerically and analytically, the system crosses the zero effective mass point very quickly. Then, the decaying oscillations with amplitude proportional to $a^{-1}$ of the effective mass render the non-relativistic Fermi gas energy density to decay, on average, as radiation, i.e. as $a^{-4}$. Similar results apply to the energy density corresponding to the scalar field. On the other hand, if the Fermi gas is initially relativistic it stays relativistic under certain initial conditions \ref{eq:conditionrel}. The scalar field eventually reaches the saturation point and oscillates. Its energy density then decays as radiation. If the condition \ref{eq:conditionrel} is not met, then the scalar field saturates after the Fermi gas reaches the non-relativistic regime. Only if we start with exactly vanishing effective mass and constant scalar field, the system remains so indefinitely. In this way, we conclude that the system always reaches a so-called scaling solution where all the components scale in the same way. In other words, all the components energy densities behave as radiation, on average. The scaling solution is broken once the scalar field mass dominates. We studied the possibility that the fermions and the scalar field make $100\%$ of dark matter in Sec.~\ref{subsec:mass}. This imposes a direct relation between the mass of the scalar field $M$ and the mass of the fermion $m_\psi$ given by \eqref{eq:Mandm}. It also predicts that the scalar field constitutes $36\%$ of the total dark matter.

We later extended our results to the non-degenerate Fermi gas in Sec.~\ref{sec:nondeg}, which should be a more realistic situation in the early universe. We found that qualitatively the coupled dynamics of the non-degenerate gas are very similar to the degenerate case. Namely, the scalar field grows, saturates around $m_{\rm eff}\sim0$ and oscillates with an amplitude proportional to $a^{-1}$. Then, the system achieves the scaling regime where all components redshift as radiation. Also, if condition \eqref{eq:conditionrel} is satisfied, an initially relativistic Fermi gas stays relativistic during the massless scalar field regime. If the Fermi gas is initially non-relativistic, then it remains non-relativistic despite that the system continuously crossing the exact relativistic $m_{\rm eff}=0$ point. Although for the non-degenerate case we only studied the relativistic and non-relativistic limits analytically, we expect that our conclusions hold if one solves the system numerically, given the similarities with the degenerate case.

Before ending this paper, let us discuss the implications of our work. We see that quite generally a massless scalar field with a Yukawa interaction to fermions in a radiation dominated universe, leads to a scaling solution where the scalar field oscillates around $m_{\rm eff}=0$ and all components behave as radiation. These background behavior definitely has some impact on the growth of perturbations in the non-relativistic regime found in Ref.~\cite{Flores:2020drq}. It would also be interesting to study in more the detail the dilatonic (exponential) coupling of \cite{Amendola:2017xhl}. The main difference is that in the exponential coupling the effective mass only vanishes asymptotically. Thus, the scalar field simply grows and there are no oscillations. This also means that since the effective mass decays in such a way that the ratio $m_{\rm eff}/T$ (or $m_{\rm eff}/n_\psi^{1/3}$) stays constant then the system is indefinitely in the non-relativistic (or relativistic) regime. In this case, the number density fluctuations of the fermions simply grow \cite{Amendola:2017xhl}. We will study the full cosmological perturbations at first order in subsequent publications. Another interesting question is whether one can find analytical solutions in the case where the massless scalar field or the Fermi gas dominates the universe, like in growing neutrino cosmologies \cite{Amendola:2007yx,Wetterich:2007kr}. We leave this issue for future work.

\section*{Acknowledgments} 
We would like to thank A. Kusenko for useful discussions and comments.
G.D. would like to thank S.~Fl{\"o}rchinger, E.~Grossi and J.~Rubio for useful correspondence. G.D. as a Fellini fellow was supported by the European Union’s Horizon 2020 research and innovation programme under the Marie Sk{\l}odowska-Curie grant 
agreement No 754496. This work was supported in part by the JSPS KAKENHI Nos.~19H01895, 20H04727 and 20H05853.

\appendix

\section{Relations in the Grand Canonical Ensemble\label{app:relations}}

In this appendix we show that the thermodynamical relations \eqref{eq:relationsbg} can also be shown explicitly. To do that, we note that the pressure \eqref{eq:p} may be also written as
\begin{align}
p_\psi=\frac{2T}{(2\pi)^3}\int d^3p\left\{ \ln \left(1+e^{-\tfrac{E-\mu}{T}}\right)+\ln\left(1+e^{-\tfrac{E+\mu}{T}}\right)\right\}\,.
\end{align}
To arrive at this expression we used that
\begin{align}
\frac{p}{E(\mathbf{p},m_{\rm eff})}f(\mathbf{p},m_{\rm eff},\mu)=-T\frac{\partial}{\partial p}\ln \left(1+e^{-\tfrac{E-\mu}{T}}\right)\,.
\end{align}
This is consistent with the fact that in the grand canonical ensemble we have $p_\psi V=-\Omega(V,\mu,T)$, where $\Omega(V,\mu,T)$ is the grand canonical potential. From this it follows that
\begin{align}
 n_\psi=\left(\frac{\partial p_\psi}{\partial\mu}\right)_{T,m_{\rm eff}}\quad{,}\quad s_\psi=\left(\frac{\partial p_\psi}{\partial T}\right)_{\mu,m_{\rm eff}}=\frac{\rho_\psi+p_\psi-\mu n_\psi}{T}\quad{\rm and}\quad \left(\frac{\partial p_\psi}{\partial m_{\rm eff}}\right)_{\mu,T}=-\frac{\rho_\psi-3p_\psi}{m_{\rm eff}}\,.
\end{align}
Other useful relations are
\begin{align}
T\left(\frac{\partial n_\psi}{\partial T}\right)_{\mu,m_{\rm eff}}+\mu\left(\frac{\partial n_\psi}{\partial \mu}\right)_{T,m_{\rm eff}}+m_{\rm eff}\left(\frac{\partial n_\psi}{\partial m_{\rm eff}}\right)_{\mu,T}=3\left(\frac{\partial p_\psi}{\partial \mu}\right)_{T,m_{\rm eff}}=3n_\psi\,.
\end{align}
Similiarly we find
\begin{align}
T\left(\frac{\partial \rho_\psi}{\partial T}\right)_{\mu,m_{\rm eff}}&+\mu\left(\frac{\partial \rho_\psi}{\partial \mu}\right)_{T,m_{\rm eff}}+m_{\rm eff}\left(\frac{\partial \rho_\psi}{\partial m_{\rm eff}}\right)_{\mu,T}\nonumber\\&
=3T\left(\frac{\partial p_\psi}{\partial T}\right)_{\mu,m_{\rm eff}}+3\mu\left(\frac{\partial p_\psi}{\partial \mu}\right)_{T,m_{\rm eff}}+\rho_\psi-3p_\psi=3\rho_\psi\,,
\end{align}
and
\begin{align}
\left(\frac{\partial \rho_\psi}{\partial m_{\rm eff}}\right)_{T,\mu}&=\frac{\rho_\psi-3p_\psi}{m_{\rm eff}}-m_{\rm eff}\left(\frac{\partial n_\psi}{\partial \mu}\right)_{T,m_{\rm eff}}\,,\\
\left(\frac{\partial \rho_\psi}{\partial \mu}\right)_{T,m_{\rm eff}}&=T\left(\frac{\partial n_\psi}{\partial T}\right)_{\mu,m_{\rm eff}}+\mu\left(\frac{\partial n_\psi}{\partial \mu}\right)_{T,m_{\rm eff}}\,.
\end{align}

\section{Series expansion \label{app:series}}

In this appendix, we expand the integral \eqref{eq:p} in a Taylor series of polylogarithmic functions of order $s$, written as ${\rm Li}_s(-x)$, based on Ref.~\cite{Trautner:2016ias}. We only focus on the pressure $p_\psi$ since all the other quantities can be derived from it. After a change of variables from $p$ to $E$ in Eq.~\eqref{eq:p} and then to $\varepsilon=T(E-m_{\rm eff})$ we arrive at
\begin{align}
p_\psi=\frac{T^4}{3\pi^2}\int_0^\infty d\varepsilon \,\varepsilon^{3/2}\left(2\frac{m_{\rm eff}}{T}+\varepsilon\right)^{3/2}\left(\frac{1}{1+e^{\varepsilon+\frac{m_{\rm eff}-\mu}{T}}}+\frac{1}{1+e^{\varepsilon+\frac{m_{\rm eff}+\mu}{T}}}\right)\,.
\end{align}
We can first expand in the relativistic limit $\varepsilon\gg m_{\rm eff}/T$. The integral then gives
\begin{align}\label{eq:prel}
p_\psi&=\frac{T^4}{3\pi^2}\sum_{n=0}^{\infty}2^n \left(\frac{m_{\rm eff}}{T}\right)^n
\binom{3/2}{n}\int_0^\infty d\varepsilon \,\varepsilon^{3-n}\left(\frac{1}{1+e^{\varepsilon+\frac{m_{\rm eff}-\mu}{T}}}+\frac{1}{1+e^{\varepsilon+\frac{m_{\rm eff}+\mu}{T}}}\right)\nonumber\\
\approx&-\frac{T^4}{3\pi^2}\sum_{n=0}^{3}2^n \left(\frac{m_{\rm eff}}{T}\right)^n
\binom{3/2}{n}\Gamma[4-n]\left({\rm Li}_{4-n}\left(-e^{-\frac{m_{\rm eff}-\mu}{T}}\right)+{\rm Li}_{4-n}\left(-e^{-\frac{m_{\rm eff}+\mu}{T}}\right)\right)\,.
\end{align}
We stopped at $n=3$ since then the integrals become ill-defined and some resummation methods are required.
We then expand in the non-relativistic limit $\varepsilon\ll m_{\rm eff}/T$. The integral then reads
\begin{align}\label{eq:pnonrel}
p_\psi&=\frac{T^4}{3\pi^2}\sum_{n=0}^\infty2^{3/2-n} \left(\frac{m_{\rm eff}}{T}\right)^{3/2-n}
\binom{3/2}{n}\int_0^\infty d\varepsilon \,\varepsilon^{3/2+n}\left(\frac{1}{1+e^{\varepsilon+\frac{m_{\rm eff}-\mu}{T}}}+\frac{1}{1+e^{\varepsilon+\frac{m_{\rm eff}+\mu}{T}}}\right)\nonumber\\
=&-\frac{T^4}{3\pi^2}\sum_{n=0}^\infty2^{3/2-n} \left(\frac{m_{\rm eff}}{T}\right)^{3/2-n}
\binom{3/2}{n}\Gamma\left[{5}/{2}+n\right]\left({\rm Li}_{{5}/{2}+n}\left(-e^{-\frac{m_{\rm eff}-\mu}{T}}\right)+{\rm Li}_{{5}/{2}+n}\left(-e^{-\frac{m_{\rm eff}+\mu}{T}}\right)\right)\,.
\end{align}
Truncating the series Eqs.~\eqref{eq:prel} and \eqref{eq:pnonrel} at a given order, we have good analytical approximation for the pressure and from Appendix~\ref{app:relations} also for the energy and number densities. In this paper we only kept the next to leading order.

We also have that the asymptotic behavior of the polylogarithmic functions is given by
\begin{align}
&\lim_{z\to0}{\rm Li}_{s}\left(-e^{z}\right)\sim-e^{z}\,, \\
&\lim_{z\to\infty}{\rm Li}_{s}\left(-e^{z}\right)\sim -\frac{x^s}{\Gamma[s+1]}\,,
\end{align}
where last one applies for $s\notin\mathbb{Z}^{-}$.

\bibliography{bibliofull.bib}

\end{document}